\documentclass[a4paper,11pt]{article}
\usepackage{jheppub}
\usepackage[dvipsnames]{xcolor}
\usepackage{comment}
\usepackage{mathtools}
\usepackage{subcaption}
\usepackage{physics}
\usepackage{tikz}
\usepackage[compat=1.1.0]{tikz-feynman}

\tikzfeynmanset{warn luatex=false}
\usetikzlibrary{arrows.meta,positioning,fit}

\newcommand{\eps}{\varepsilon}
\newcommand{\rd}{\mathrm{d}}
\newcommand{\cS}{\mathcal{S}}
\newcommand{\cQ}{\mathcal{Q}}
\newcommand{\cO}{\mathcal{O}}
\newcommand{\cI}{\mathcal{I}}
\DeclareMathOperator{\Gram}{Gram}
\DeclareMathOperator{\Vol}{Vol}
\newcommand{\iu}{{i\mkern1mu}}
\DeclareMathOperator{\Li}{Li}

\DeclareMathOperator{\sgn}{sgn}
\newcommand{\Q}{\mathcal{Q}}

\def\beq{\begin{equation}}
\def\eeq{\end{equation}}
\def\bsp#1\esp{\begin{split}#1\end{split}}

\author{Paul Mork${}^1$}
\affiliation{
\vskip 0.5 em
${}^1$Bethe Center for Theoretical Physics, Universit\"at Bonn, D-53115, Germany\\
\vskip 0.5 em}

\emailAdd{pmork@uni-bonn.de}

\title{Recursive construction of scalar one-loop integrals in dimensional regularisation}

\abstract{
We derive a novel recursive structure for dimensionally regularised scalar one-loop Feynman integrals based on Schläfli's differential formula for hyperbolic simplices. The recursion relates the Laurent coefficients in the dimensional regulator \(\eps\) of an \(N\)-point integral to lower-order coefficients of integrals with additional external legs. 
The construction is seeded by the \(\eps = 0\) contributions, which admit a geometric interpretation as volumes of simplices in hyperbolic space and are known in terms of multiple polylogarithms (MPLs).
Iterating the recursion therefore provides a constructive algorithm for computing arbitrary orders in the \(\eps\)-expansion of scalar one-loop integrals with arbitrary masses and kinematics, while remaining entirely within the class of MPLs.
In particular, this establishes that all coefficients in the Laurent expansion of dimensionally regularised scalar one-loop integrals can be expressed in terms of MPLs. 
As a first application beyond existing results, we obtain an explicit closed MPL expression for the \(\mathcal O(\eps)\) coefficient of the scalar hexagon with arbitrary masses and off-shell Euclidean kinematics.
}

\begin{document}
\begin{flushright}
    BONN-TH/2026-15
    \end{flushright}

\maketitle
\flushbottom

\section{Introduction}
Precise predictions for collider experiments require quantitative control over radiative corrections to scattering amplitudes at increasingly high orders in perturbation theory (see,~e.g., refs.~\cite{Heinrich:2020ybq,Caola:2022ayt}). Over the past decades, remarkable progress has been achieved in the computation of multi-loop amplitudes~\cite{Henn:2013pwa,Duhr:2014woa}, including two-, three-, and even higher-loop contributions~\cite{Dixon:2013eka}. These developments have not only led to powerful computational techniques~\cite{Chetyrkin:1981qh,Laporta:2000dsw,Goncharov:2010jf}, but have also revealed deep connections to areas of mathematics such as algebraic geometry and number theory~\cite{Brown:2015fyf,Schnetz:2013hqa}, turning the study of scattering amplitudes into a highly active and interdisciplinary field of research.

In comparison, one-loop integrals are often regarded as a well-understood subject~\cite{tHooft:1978jhc,Passarino:1978jh,Davydychev:1997wa,Schnetz:2010pd}, at least in the case of integer space-time dimensions. 
In this setting, scalar one-loop integrals are closely related to volumes of simplices in hyperbolic space~\cite{Davydychev:1997wa,Schnetz:2010pd,Mason:2010pg,Nandan:2013ip}, which can be computed systematically in terms of multiple polylogarithms (MPLs)~\cite{Goncharov:1998kja,Goncharov1999Volumes,Rudenko2020Orthoschemes,Ren:2023tuj}. 
As a consequence, scalar one-loop integrals at \(\eps = 0\) admit representations in terms of MPLs for arbitrary kinematics and any number of external legs.

The situation is markedly different for dimensionally regulated Feynman integrals~\cite{tHooft:1972tcz,Cicuta:1972jf,Bollini:1972ui}. 
Extending the analytic control beyond \(\eps = 0\) is highly non-trivial, as the complexity increases with the transcendental weight at higher orders in \(\eps\). 
Indeed, explicit results are only known in a limited number of cases with few external legs (see, e.g.,~refs.~\cite{DelDuca:2009ac,Dixon:2011ng,DelDuca:2011ne,DelDuca:2011jm,DelDuca:2011wh,Chavez:2012kn,Kozlov:2015kol,Bourjaily:2019exo,Haug:2022hkr,Haug:2023eqg,Buccioni:2023okz,Becchetti:2025osw}).

Nevertheless, understanding higher orders in the dimensional regulator \(\eps\) is required in a variety of contexts. 
At higher loop orders, for example, products of lower-loop integrals containing poles in \(\eps\) can combine with positive powers of \(\eps\) from other contributions to yield finite terms in scattering amplitudes. 
This makes the computation of the Laurent expansion of one-loop integrals beyond leading order essential, even for observables at higher perturbative orders.

In ref.~\cite{Duhr:2025azh}, an approach based on direct integration techniques~\cite{Brown2009MasslessHigherLoop,Anastasiou:2013srw,Ablinger:2014yaa,Panzer:2014caa,hyperlogprocedures,Kardos:2025klp} was proposed, yielding explicit results for triangle, box, and pentagon integrals to arbitrary order in \(\eps\). 
A key step in that construction was the rationalisation of square roots~\cite{Besier:2018jen,Besier:2019kco} appearing in the MPL arguments. 
Starting from the hexagon, however, scalar one-loop integrals generically involve several independent square roots that cannot be rationalised simultaneously, and no extension of this method to higher-point integrals has been achieved so far.

This motivates the search for alternative approaches that avoid direct integration and rationalisation.
In this paper we overcome this obstruction by adopting a different strategy. Instead of performing direct integrations of MPLs, we exploit Schläfli's differential formula~\cite{Schlafli1901} for hyperbolic simplices. Using the dissection of hyperbolic simplices into orthoschemes from ref.~\cite{Duhr:2025azh}, we derive a recursive relation that connects the \(\eps^k\) coefficient of an \(N\)-point function (\(N\) even) to the \(\eps^{k-1}\) coefficient of an \((N+1)\)-point function, thereby increasing the number of legs while lowering the order in \(\eps\). The remaining cases with odd \(N\) are obtained from even \((N+1)\)-point functions through a limiting procedure (sending a dual point to infinity), in line with the extended dual conformal invariance of the one-loop integrals~\cite{Henn:2011by,Loebbert:2020hxk,Loebbert:2020glj}. The recursion is seeded by the \(\eps = 0\) results, which are known in terms of MPLs. In this way, the computation of higher orders in \(\eps\) is reduced to evaluating controlled kinematic limits of higher-point integrals. 
In total, this provides a closed recursive framework that determines the Laurent expansion of scalar one-loop integrals in dimensional regularisation to any desired order in \(\eps\), while remaining entirely within the class of multiple polylogarithms, i.e.\ no functions beyond MPLs are generated at any stage of the recursion.


%
As a first application of the method beyond existing results, we compute the \(\mathcal{O}(\eps)\) coefficient of the scalar hexagon integral, which was previously out of reach of existing approaches. The result is expressed entirely in terms of MPLs with comparatively simple arguments. 

The paper is organised as follows. 
In section~\ref{sec:setup} we introduce dimensionally regularised one-loop integrals, including their auxiliary mass representation and Laurent expansion, and recall their relation to hyperbolic simplices. 
We also discuss their dissection into orthoschemes, which forms the basis of our recursive construction.
Section~\ref{sec:schlaefli} contains our main result, the $\eps$-recursion. 
After reviewing Schläfli's differential equation in section~\ref{sec:SchlaefliFormula}, we derive the first-order relations in section~\ref{sec:schlaefli_rec-1} and generalise them to all orders in section~\ref{sec:AllOrdersRecurrence}. 
In section~\ref{sec:consequences} we discuss structural implications of the recursion, including that all Laurent coefficients in \(\eps\) of one-loop integrals can be expressed through MPLs (section~\ref{sec:closure}) and the associated differential equation, reproducing the \(\mathcal{O}(\eps)\) differential equation of ref.~\cite{Abreu:2017mtm} (section~\ref{app:deq}). 
Section~\ref{app:examples} presents explicit examples, including bubble, triangle, and box integrals. Therein, we present a geometric interpretation of the \(\mathcal{O}(\eps)\) term of the bubble diagram through a volume in Anti-de-Sitter space.
As a first non-trivial application of the method, the hexagon at $\mathcal{O}(\eps)$ is treated separately in section~\ref{sec:hexagon}.
We conclude in section~\ref{sec:conclusions}. 
The paper is supplemented by appendix~\ref{app:evenodd}, which contains the proof of the even-to-odd conversion~\eqref{eq:eventoodd}.
\section{Setup}\label{sec:setup}
In this section we introduce the notation and conventions for one-loop scalar integrals that will be used throughout the paper. We begin by defining the class of integrals under consideration.

\subsection{Definition of one-loop integrals}
We study one-loop scalar integrals with $N$ propagators and external momenta $p_i$, which obey overall momentum conservation, \(\sum_{i=1}^N p_i = 0 \).
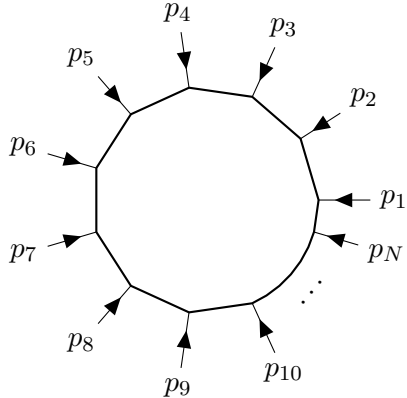
\begin{figure}[htbp]
    \centering
    \begin{tikzpicture}
    \begin{feynman}
        \def\n{11}
        \def\r{1.5}

        \foreach \i in {1,...,10} {
            \pgfmathsetmacro{\angle}{360/\n*(\i-1)}
            \pgfmathsetmacro{\x}{\r*cos(\angle)}
            \pgfmathsetmacro{\y}{\r*sin(\angle)}
            \vertex (v\i) at (\x,\y);

            \pgfmathsetmacro{\xo}{(\r+1)*cos(\angle)}
            \pgfmathsetmacro{\yo}{(\r+1)*sin(\angle)}
            \vertex (p\i) at (\xo,\yo) {\(p_{\i}\)};
            
            \diagram* {
                (p\i) -- [fermion] (v\i),
            };
        }

        \foreach \i in {11,...,15} {
            \pgfmathsetmacro{\angle}{360/\n/6*1.5*(\i-10)+360/\n*(9)}
            \pgfmathsetmacro{\x}{\r*cos(\angle)}
            \pgfmathsetmacro{\y}{\r*sin(\angle)}
            \vertex (v\i) at (\x,\y);

            \pgfmathsetmacro{\xo}{(\r+1)*cos(\angle)}
            \pgfmathsetmacro{\yo}{(\r+1)*sin(\angle)}
            \vertex (p\i) at (\xo,\yo) {};
        }
        \pgfmathsetmacro{\angle}{360/\n*(10.5)}
        \pgfmathsetmacro{\x}{\r*cos(\angle)}
        \pgfmathsetmacro{\y}{\r*sin(\angle)}
        \vertex (v16) at (\x,\y);

        \pgfmathsetmacro{\xo}{(\r+1)*cos(\angle)}
        \pgfmathsetmacro{\yo}{(\r+1)*sin(\angle)}
        \vertex (pn) at (\xo,\yo) {\(p_N\)};

        \diagram* {
            (pn) -- [fermion] (v16),
        };

        \draw[thick] (v1) 
            \foreach \i [evaluate={\j=int(mod(\i,16)+1)}] in {1,...,15} 
                { -- (v\j) } -- (v16) -- cycle;

        \pgfmathsetmacro{\angleMid}{360/\n*9.75}
        \pgfmathsetmacro{\angleMidd}{\angleMid+90}
        \pgfmathsetmacro{\dx}{\r*1.25*cos(\angleMid)}
        \pgfmathsetmacro{\dy}{\r*1.25*sin(\angleMid)}
        \node[rotate=\angleMidd] at (\dx,\dy) {\(\cdots\)};
    \end{feynman}
    \end{tikzpicture}
    \caption{Schematic depiction of a one-loop \(N\)-point Feynman graph}
    \label{fig:n-gon_graph}
\end{figure}\\
Throughout this work we employ dimensional regularisation~\cite{tHooft:1972tcz,Cicuta:1972jf,Bollini:1972ui} in \(D=d-2\eps\) dimensions, with \(d\) an even integer. In dimensional regularisation and Euclidean kinematics, these integrals are defined as
\begin{equation}\label{eq:integral_def}
I_N^D(\{p_i\},\{m_i\})
= e^{\gamma_\textnormal{E}\eps}
\int \frac{\mathrm{d}^D l}{\pi^{D/2}}
\prod_{i=1}^N
\frac{1}{(l-l_i)^2 + m_i^2} \, ,
\end{equation}
where \(m_i\) denote the masses of the internal propagators and \(l_i = \sum_{j=1}^{i-1} p_j\) are partial sums of the external momenta. The constant \(\gamma_\textnormal{E}=-\Gamma'(1)\) is the Euler–Mascheroni constant.
Instead of resorting to the masses and momenta as the arguments of eq.~\eqref{eq:integral_def} it will prove useful to encode all the kinematic dependence in the \(N \times N\) Cayley matrix:
\beq\label{eq:Cayley}
Q_{i,j} = \frac{(l_i - l_j)^2 + m_i^2 + m_j^2}{2}\,.
\eeq
These entries are the coefficients of the Symanzik \(\mathcal{F}\)-polynomial. We will from now on write \(I_N^D(Q)\) instead of \(I_N^D(\{p_i\},\{m_i\})\).

Using integration-by-parts relations~\cite{Tkachov:1981wb,Chetyrkin:1981qh} and tensor reduction~\cite{tHooft:1978jhc,Passarino:1978jh}, general one-loop integrals can be reduced to scalar integrals of the form above.
Moreover, dimensional recurrence relations~\cite{Tarasov:1996bz,babis_thesis,Lee:2009dh} allow one to restrict the integer dimension to \(d=N\) or \(d=N+1\).

In strictly four space-time dimensions, one-loop amplitudes reduce to integrals with at most four propagators~\cite{Melrose:1965kb}. In dimensional regularisation, pentagon integrals may additionally contribute through terms suppressed by powers of \(\eps\). From a phenomenological perspective, this indicates that integrals with \(N\le 5\) are sufficient for fixed-order computations in \(D=4-2\eps\).

Nevertheless, integrals with arbitrary numbers of propagators are of intrinsic interest beyond phenomenology. A formulation valid for general \(N\) may uncover structural features that are obscured at low multiplicity. Moreover, although one-loop integrals in integer dimensions \(d=N\) admit a geometric interpretation in terms of hyperbolic simplices~\cite{Davydychev:1997wa,Ren:2023tuj,Duhr:2025azh}, it remains unclear how this geometric picture extends to dimensional regularisation.

In the remainder of this section, we review the necessary tools, i.e., the auxiliary mass integration representation, the relationship between one-loop integrals in integer dimension and hyperbolic simplices, as well as dissection techniques.

\subsection{Auxiliary mass integration}\label{sec:auxiliary-mass}

The goal of this section is to reformulate the dimensionally regularised Feynman integral \(I_N^D(Q)\), defined in eq.~\eqref{eq:integral_def}, as a one-dimensional integral over a one-parameter family of integrals \(I_N^d(Q_\eta)\) in integer space-time dimension. 
In this representation, the dependence on the dimensional regulator \(\eps\) is entirely captured by the integration measure.

Following ref.~\cite{Duhr:2025azh}, this reduces the problem of determining higher orders in \(\eps\) to analysing a one-parameter deformation of integer-dimensional integrals. 
A related dimension-changing representation was used for high-order numerical evaluations of one-loop integrals in ref.~\cite{Huang:2024dct}.
This representation makes it straightforward to relate the integral to Schläfli's differential equation for hyperbolic volumes, which forms the basis of our recursive construction in section~\ref{sec:schlaefli}.

In the following, we restrict to cases where the integer-dimensional integral \(I_N^d(Q_\eta)\) is finite, which suffices for the class of integrals relevant to our construction.\footnote{This excludes certain degenerate low-point configurations, such as special bubble integrals, which require a separate treatment. These integrals are, however, all known analytically.}
We work in the ’t~Hooft–Veltman scheme, where the external momenta lie in \(d\) dimensions, while the loop momentum \(l\) is taken in \(D=d-2\eps\) dimensions. We split the loop momentum into its \(d\)-dimensional component and its \((-2\eps)\)-dimensional component,
\beq
l = l_{\parallel} + l_{\bot}\,,
\eeq
with \(l_{\parallel}\cdot l_i = l\cdot l_i\) and \(l_{\bot}\cdot l_i = 0\). Writing
\beq
\eta = l_{\bot}^2\,,
\eeq
the propagators take the form
\beq
(l-l_i)^2 + m_i^2
= (l_{\parallel}-l_i)^2 + m_i^2 + \eta\,.
\eeq
Introducing spherical coordinates in the \((-2\eps)\)-dimensional subspace, the integration measure factorises as
\beq
\mathrm{d}^D l
= \frac{1}{2}\,\eta^{-1-\eps}\,
\mathrm{d}^d l_{\parallel}\,
\mathrm{d}\Omega_{-2\eps}\,
\mathrm{d}\eta\,,
\eeq
where \(\Omega_{-2\eps}\) denotes the \((-2\eps)\)-dimensional solid angle.
Performing the angular integration yields
\begin{equation}\label{eq:auxmass_master}
I_N^{d-2\eps}(Q)
=
\frac{e^{\gamma_\textnormal{E}\eps}}{\Gamma(-\eps)}
\int_0^\infty
\mathrm{d}\eta\,
\eta^{-1-\eps}\,
I_N^d(Q_\eta)\, ,
\end{equation}
where \(Q_\eta\) is obtained from \(Q\) by shifting all squared masses according to
\beq
m_i^2 \to m_i^2 + \eta\, .
\eeq
For this reason we refer to eq.~\eqref{eq:auxmass_master} as the \emph{auxiliary mass integration}.
From the definition of the Cayley matrix in
eq.~\eqref{eq:Cayley} it is clear that 
\beq Q_\eta = Q + \eta \, \mathbf{1}\mathbf{1}^\top \, ,\eeq
where \(\mathbf{1}\) is a \(N\)-dimensional column vector whose entries are all one. The parameter \(\eta\) thus generates a rank-one deformation \(Q \to Q_\eta\) of the Cayley matrix \(Q\).

\subsubsection{Laurent expansion}
We are interested in the \(\eps\)-expansion of eq.~\eqref{eq:auxmass_master}. Since the integral 
\beq\label{eq:aux_mass_integral_part} \mathcal{I}_{N,d} (Q, \eps) \coloneqq \int_0^\infty \mathrm{d}\eta\,\eta^{-1-\eps}\,I_N^d(Q_\eta)\eeq
 has a simple pole at \(\eps=0\) that gets cancelled by the prefactor 
 \beq \frac{e^{\gamma_\textnormal{E} \eps}}{\Gamma (-\eps)} = -\eps + \mathcal{O}(\eps^2) \, , \eeq %
we can not simply expand eq.~\eqref{eq:aux_mass_integral_part} under the integral sign.
Another source of divergence could be the boundary at infinity. However, for large \(\eta\), the integer-dimensional integral decreases as a negative power of \(\eta\) (by power counting in the large-mass limit), so that no divergence arises from the upper integration boundary.

To isolate the singularity of the integral as \(\eps \to 0 \), we first perform the change of variables \(\eta = \tfrac{u}{1-u}\). Under this transformation the differential becomes:
\begin{equation}
    \frac{\mathrm{d}\eta}{\eta} = \mathrm{d}u \left( \frac{1}{u} + \frac{1}{1-u} \right)\, .
\end{equation}
The pole at \(u=0\) corresponds to the pole at \(\eta = 0\) while the \(u=1\) is the spurious pole at \(\eta=\infty\). 

For simplicity of notation 
we denote \(I(u) \coloneqq I_N^d (Q_{u/(1-u)})\). The change of coordinates then yields
in eq.~\eqref{eq:aux_mass_integral_part} leaves us with
\beq\label{eq:aux_mass_integral_part_u}
    \mathcal{I}_{N,d} (Q, \eps) = \int_0^1 \mathrm{d}u \, u^{-1-\eps} \, (1-u)^\eps I(u) + \int_0^1 \mathrm{d}u \, \frac{1}{1-u} \, u^{-\eps} \, (1-u)^\eps I(u)\, ,
\eeq
where the second part involving \(\tfrac{1}{1-u} \) does not generate a pole in \(\eps\), since the integrand is integrable at \(u=1\). The first term, however, generates a pole in \(\eps\) and requires further treatment.
This can be achieved by using the distributional expansion of \(u^{-1-\eps}\):
\beq
u^{-1-\eps} = -\frac{1}{\eps}\,\delta(u) + \sum_{k=0}^{\infty}\frac{(-\eps)^k}{k!} \,\left[\frac{(\log u )^k}{u}\right]_+\,. 
\eeq
The plus-distribution \(\left[\tfrac{(\log u )^k}{u}\right]_+ \) is defined by its action on a test function \(\varphi (x)\), i.e.,
\beq
\int_0^1\rd u\,\varphi(u)\,\left[\frac{\log^k u}{ u }\right]_+ := \int_0^1\rd u\,\frac{[\varphi(u )-\varphi(0)]\,\log^k u }{ u}\, .
\eeq
Then the first summand in eq.~\eqref{eq:aux_mass_integral_part_u} can be rewritten as:
\begin{equation}\label{eq:IetaIntegral}
    \int_0^1 \mathrm{d}u \, u^{-1-\eps} \, (1-u)^\eps I(u) = - \frac{1}{\eps} I(0) + \sum_{k=0}^\infty \frac{\eps^k}{k!}\, \int_0^1 \mathrm{d}{u}\, \frac{\left[ (1-u)^\eps I(u) -I(0) \right] (- \log u )^k }{u} \, .
\end{equation}
This isolates the pole in \(\eps\), i.e., \(-\tfrac{1}{\eps} I(0)\), from the remainder of the integral, which is regular. Note that \(I(0) = I_N^d (Q)\).
Combining everything and 
expanding the remaining \(\eps\)-dependent factors in the integrand around \(\eps=0\) yields a Laurent expansion
\beq\label{eq:Laurent_Expansion}
    \mathcal{I}_{N,d} (Q, \eps) = \sum_{k=-1}^\infty \mathcal{I}_{N,d}^k (Q) \eps^k \, ,
\eeq
with Laurent coefficients
\beq
    \mathcal{I}_{N,d}^{-1} = - I_N^d(Q)\, ,
\eeq
and for \(k \ge 0\)
\begin{equation}\label{eq:eps_expansion_coeff}
\begin{split}
    \mathcal{I}_{N,d}^k (Q) \coloneqq \frac{(-1)^k}{k!} \bigg( &\int_0^1 \frac{\mathrm{d} u}{u} \left( \log \left( \frac{u}{1-u} \right)^k  I (u) - \log (u)^k I_N^{d} (Q) \right) \\
    &\qquad + \int_0^1 \frac{\mathrm{d} u }{1-u} \log \left( \frac{u}{1-u} \right)^k I(u) \bigg)\, .
\end{split}
\end{equation}
We might write \(\mathcal{I}_{N}^k (Q) = \mathcal{I}_{N,N}^k (Q)\) when there is no confusion. Note that there is a shift of \(+1\) speaking about the \(\cO(\eps)\) term of an \(N\)-point function and the definition of Laurent coefficients in eq.~\eqref{eq:Laurent_Expansion} since the Laurent series misses the prefactor \(\tfrac{e^{\gamma_\mathrm{E} \eps}}{\Gamma (-\eps)} = - \eps + \cO(\eps^2)\) and thus starts at \(\tfrac{1}{\eps}\).

Since the \(\eps=0\) contributions are known in terms of multiple polylogarithms (MPLs) (see section~\ref{sec:hyperbolic_volume}), it is natural to investigate whether the higher-order Laurent coefficients can also be described within this function space. We therefore briefly review the relevant properties of MPLs in the following.
MPLs may be defined recursively as iterated integrals~\cite{Goncharov:1998kja}:
\begin{equation}
G(a_1,\ldots,a_n; x)
= \int_0^x \frac{\mathrm{d}t}{t-a_1} G(a_2,\ldots,a_n; t)\,,
\end{equation}
with the condition that:
\beq
G(\underbrace{0,\ldots,0}_{n};x)
=
\frac{1}{n!}\log^n x\,.
\eeq
The number of integrations \(n\) is called the \emph{weight} of the MPL. Accordingly, we define
\beq
\operatorname{w}\!\left(G(a_1,\ldots,a_n;x)\right)=n \, .
\eeq
More generally, a linear combination of MPLs is said to have weight \(n\) if each term has weight \(n\). By a \emph{pure MPL} expression we mean a linear combination of multiple polylogarithms without additional algebraic prefactors.

Another representation of MPLs is as nested sums, i.e.,
\beq\label{eq:Li_def}
\Li_{m_1,\ldots,m_k}\!(z_1,\ldots,z_k) = \sum_{0<n_1<\cdots<n_k}\frac{z_1^{n_1}}{n_1^{m_1}}\cdots\frac{z_k^{n_k}}{n_k^{m_k}}\,.
\eeq
This representation is related to the iterated integral form via
\beq
\Li_{m_1,\ldots,m_k}(z_1,\ldots,z_k)
=
(-1)^k\,
G(\underbrace{0,\ldots,0}_{m_k-1},\tfrac{1}{z_k},\ldots,
\underbrace{0,\ldots,0}_{m_1-1},\tfrac{1}{z_1\cdots z_k};1)\,.
\eeq
In particular, classical polylogarithms \(\Li_n(x)\) arise as special cases.
We will encounter MPLs more explicitly in the following sections, where they arise from the evaluation of the integer-dimensional integrals \(I_N^d(Q)\).

The key advantage of the auxiliary mass integration is that all higher-order terms in the Laurent expansion are expressed as one-dimensional integrals over the corresponding integer-dimensional Feynman integral, multiplied by (powers of) logarithms with simple arguments. In particular, eq.~\eqref{eq:eps_expansion_coeff} implies that if \(\mathcal{I}_{N,d}^k (Q)\) can be expressed through MPLs then its weight should equal:
\beq\label{eq:weight_counting}
    \operatorname{w}\!\left(\mathcal{I}_{N,d}^k (Q)\right) = \frac{d}{2} + k + 1\, .
\eeq

In the next section we review the relation between one-loop integrals in integer space-time dimension and volumes of hyperbolic simplices, which provides an explicit representation of the integrand \(I_N^d (Q)\) appearing in the auxiliary mass formula~\eqref{eq:auxmass_master}.

\subsection{Connection to volumes in hyperbolic geometry}\label{sec:hyperbolic_volume}
We now recall that one-loop integrals in integer dimension \(D=N\) (even) admit a geometric interpretation in terms of hyperbolic volumes~\cite{Davydychev:1997wa,Ren:2023tuj,Duhr:2025azh}.
We start by rewriting eq.~\eqref{eq:integral_def} in the Feynman-parameter representation:
\begin{equation}\label{eq:I_N^N_integral}
    I_N^N (Q) = \Gamma\! \left( \tfrac{N}{2} \right) \int_0^\infty \frac{\delta(\alpha_N - 1)\,\prod_{i=1}^N \mathrm{d} \alpha_i }{\left(\sum_{i,j=1}^N Q_{i,j}\, \alpha_i\, \alpha_j\right)^{N/2}}\,,
\end{equation}
where all the kinematic dependence is encoded into the Cayley matrix (see eq.~\eqref{eq:Cayley}). 
This expression closely resembles the integral representation of the volume of simplices in hyperbolic \(\mathbb{H}^{N-1}\), as we now recall.
Hyperbolic space \(\mathbb H^{N-1}\) can be realised in the projective (Klein) model as the unit ball in \(\mathbb R^{N-1}\), equipped with the bilinear form
\beq\label{eq:Scalar_Product_Klein}
\langle p,q\rangle = 1 - \sum_{k=1}^{N-1} p_k q_k \, .
\eeq
Given \(N\) points \(v_1,\ldots,v_N \in \mathbb H^{N-1}\), their convex hull defines a hyperbolic simplex \(\mathcal S\) with Gram matrix defined by
\beq
\Gram(\mathcal S) = \big(\langle v_i,v_j\rangle\big)_{1\le i,j\le N}\, .
\eeq
The volume of the simplex \(\cS\) depends only on its Gram matrix (or, equivalently, on its dihedral angles). 
Its variations under deformations are governed by Schläfli's differential equation~\cite{Schlafli1901}. We will return to this fundamental relation in the main part (section~\ref{sec:schlaefli}) of the paper, where it provides the mechanism underlying our construction.

An explicit integral representation of the volume of the simplex \(\cS\) is given by:
\beq\label{eq:volume_general}
\Vol(\cS) = \sqrt{|\det \Gram(\cS)|}\,  
\int_0^\infty 
\frac{\delta(\alpha_N - 1)\,\prod_{i=1}^N \mathrm{d} \alpha_i}
{\left(\sum_{i,j=1}^N  \langle v_i, v_j\rangle\,\alpha_i\, \alpha_j\right)^{N/2}}\,.
\eeq

Upon comparison with the Feynman-parameter representation in eq.~\eqref{eq:I_N^N_integral} we find
\begin{equation}\label{eq:INN_Vol}
    I_N^N (Q) = \Gamma\! \left( \tfrac{N}{2} \right) 
    \frac{\operatorname{Vol}(\cS_Q)}{\sqrt{|\det Q|}}\,,
\end{equation}
where \(\cS_Q\) denotes a hyperbolic simplex whose Gram matrix satisfies
\beq
\Gram (\cS_Q) = Q \, .
\eeq
Strictly speaking, the Cayley matrix \(Q\) does not directly coincide with the Gram matrix of a hyperbolic simplex in the standard normalisation. However, owing to the invariance of eq.~\eqref{eq:volume_general} under non-vanishing rescalings \(Q_{ij} \to a_i a_j Q_{ij}\), it can always be brought to such a form. 
In Euclidean kinematics, the matrix \(Q\) has Lorentzian signature \((1,N-1)\), which implies that it can be realised as the Gram matrix of a hyperbolic simplex \(\cS_Q\).\footnote{Assuming generic Euclidean kinematics, restricting to the subspace $\sum_i x_i = 0$ one finds $x^\top Q x = -(\sum_i x_i l_i)^2 < 0$, yielding $N-1$ negative eigenvalues. Since $Q_{ii}=m_i^2>0$, there is also a positive direction.}
Thus far, eq.~\eqref{eq:INN_Vol} reduces the computation of \(I_N^N(Q)\) to the evaluation of the hyperbolic volume \(\Vol(\mathcal S_Q)\).

We have restricted ourselves to the case where \(N\) is even. This is not a limitation, since for odd \(N\) (corresponding to \(D = N+1\) in our setup), the integral can be related to an \((N+1)\)-point integral in \(D = N+1\) with one vertex sent to infinity.
Geometrically, this amounts to working with the \emph{extended Gram matrix} \(Q_{\mathrm{ext}}\), defined as the \((N+1)\times(N+1)\) matrix
\begin{equation}\label{eq:extended_Gram}
    Q_{\mathrm{ext}} \coloneqq  
        \begin{pmatrix}
        Q & \tfrac{1}{2}\mathbf{1} \\
        \tfrac{1}{2}\mathbf{1}^\top & 0
        \end{pmatrix}.
\end{equation}
With this extension one has\footnote{The vanishing last diagonal entry in \(Q_{\mathrm{ext}} \) may lead to spurious divergences, in which case it is replaced by a regulator $\delta>0$, and the limit $\delta \to 0$ is taken at the end.}
\begin{equation}
    I_{N}^{N+1}(Q) = I_{N+1}^{N+1}(Q_{\mathrm{ext}})\,,
\end{equation}
so that also in this case the integral is represented by the volume of a hyperbolic simplex, now with one ideal vertex.
Hence the geometric interpretation developed above applies uniformly to both cases \(D=N\) and \(D=N+1\). Henceforth, we will with slight abuse of notation write \(\cS_Q\) instead of \(\cS_{Q_\textnormal{ext}}\) and also call \(Q\) the Gram matrix when we actually mean \(Q_\textnormal{ext}\). It should always be clear from context and the dimension of \(Q\) which matrix to use.

The problem of determining volumes of hyperbolic simplices has a long history, beginning with the foundational work of Lobachevsky~\cite{Lobachevsky1837}. Subsequent developments in special functions and hyperbolic geometry further advanced the subject~\cite{coxeter_functions_1935,Bhm1964ZuCI,wagner_volume_1996,Kellerhals1995Volumes,Cho1999OnTV,Goncharov1999Volumes,Murakami2005OnTV}. 
Among hyperbolic simplices, a particularly important class is given by \emph{orthoschemes}. An orthoscheme can be regarded as the higher-dimensional analogue of a right-angled triangle: it is defined by an ordered set of hyperplanes \(H=(H_1,\ldots,H_N)\) in \(\mathbb H^{N-1}\) in general position such that \(H_i \perp H_j\) whenever \(|i-j|>1\).

Building on the classical theory and on Goncharov's framework relating hyperbolic volumes and polylogarithms, Rudenko~\cite{Rudenko2020Orthoschemes} obtained an explicit formula for the volume of hyperbolic orthoschemes in terms of MPLs (for odd-dimensional hyperbolic space).

Every hyperbolic simplex can be decomposed algorithmically into a finite collection of orthoschemes~\cite{Davydychev:1997wa,Duhr:2025azh}. In general this dissection is signed, so that the total volume is obtained as a signed sum of orthoscheme volumes (see section~\ref{sec:dissections}). As a consequence, the volume \(\Vol(\mathcal S_Q)\) admits a representation in terms of MPLs in Euclidean kinematics.

Since the present work focuses on structural properties rather than explicit closed-form expressions, we will not require the detailed form of Rudenko's formula. It suffices for our purposes that integer-dimensional one-loop integrals \(I_N^N(Q)\) can be expressed in terms of MPLs. Explicit representations can be found in~\cite{Rudenko2020Orthoschemes,Ren:2023tuj,Duhr:2025azh}.

\subsection{Dissecting the integrals}\label{sec:dissections}
Since an explicit closed formula for \(\Vol(\cS_Q)\) is not available in general, the simplex \(\cS_Q\) needs to be decomposed into orthoschemes, i.e.,
\beq
    \cS_Q = \sum_{\cQ} \sgn (\cS_{\cQ}) \, \cS_{\cQ}\, .
\eeq 
Here the simplices \(\cS_{\cQ}\) are orthoschemes and \(\sgn (\cS_{\cQ})\) is an appropriate sign to be defined later in eq.~\eqref{eq:sgn_Factor} for our choice of dissection. For \(\Vol (\cS_{\cQ} ) \) an explicit formula in terms of MPLs exists as discussed above~\cite{Rudenko2020Orthoschemes}. 
Moreover, this decomposition drastically reduces the number of independent kinematic parameters for each integral: while \(I_N^D(Q)\) depends on up to \(\tfrac{N}{2}(N+1)\) scales, each orthoscheme involves at most \(N\).

However, one issue with this decomposition is that it is not a priori clear whether it extends to dimensionally-regularised Feynman integrals, rather than only to volumes (\(\eps=0\)).
In ref.~\cite{Duhr:2025azh}, it was shown that this holds for a special dissection known as the \emph{projective dissection}, which is equivalent to the so-called \emph{splitting of the basic simplex} in ref.~\cite{Davydychev:1997wa}. We will refer to this dissection through its set of Gram matrices \(\cQ \in \operatorname{BS}(Q)\).

We now describe the elements of \(\operatorname{BS}(Q)\).
It is convenient to label its elements by an index sequence \(\mathbf{j}=(j_1,\ldots,j_N)\) of \(\{1, \ldots, N\}\). The entries of \(\cQ(\mathbf{j})\) are then given by (\(i \le k\)) \cite[eq.~3.8]{Duhr:2025azh}:
\beq\label{eq:mathcalQ_coords}
    \cQ_{i,k}(\mathbf{j}) = \cQ_{k,k}(\mathbf{j}) = \frac{\det Q_{\mathbf{j}_{[k]}}}{-4 \det Q_{\mathbf{j}_{[k]}}^\textnormal{ext}}\, ,
\eeq
where \(\cQ_{I} \coloneqq \cQ_{I,I}\) with \(\cQ_{I,J}\) being the submatrix of \(\cQ\) formed by rows indexed by \(I\) and columns indexed by \(J\), \(Q_{I}^\textnormal{ext}= \begin{pmatrix}
    \cQ_I & \tfrac{1}{2}\, \mathbf{1} \\ \tfrac{1}{2}\, \mathbf{1}^\top & 0
\end{pmatrix}\), and \(\mathbf{j}_{[k]}=(j_1,j_2,\ldots,j_k)\) denotes the restriction of \(\mathbf{j}\) to the first \(k\) elements.
The sign factor is given by \cite[eq.~3.12]{Duhr:2025azh}
\begin{equation}\label{eq:sgn_Factor}
    \operatorname{sgn}(\cS_{\cQ (\mathbf{j})})=\sgn(\mathbf{j})\prod_{i=1}^{N}\sgn(\mathbf{j}_{[i]}) \, \sgn \left( \frac{\det
    \begin{pmatrix}
        Q_{\mathbf{j}_{[i]},\mathbf{j}_{[i-1]}} & \frac{1}{2}\mathbf{1}
    \end{pmatrix}
    }{-4 \det Q^\textnormal{ext}_{\mathbf{j}_{[i]}}} \right)\, .
\end{equation}
We often omit the specific index sequence \(\mathbf{j}\) when there is no need to differentiate between different orthoschemes of \(\operatorname{BS}(Q)\). In practice we only need to restore \(\mathbf{j}\) in the end when we sum orthoscheme contributions.

In particular, in ref.~\cite{Duhr:2025azh} it was shown that this dissection procedure into \(\operatorname{BS}(Q)\) is compatible with the auxiliary mass integration deformation of the Gram matrix in the sense that for \(\cQ' \in \operatorname{BS}(Q_\eta) \) there exists a unique element \(\cQ \in \operatorname{BS}(Q)\) such that \(\cQ' = \cQ_\eta = \cQ + \eta \, \mathbf{1}\mathbf{1}^\top\), i.e., dissecting the deformed Gram matrix \(Q_\eta\) simply amounts to deforming the orthoschemes in the dissection of \(Q\). 
The dimensionally-regularised integral can thus be brought into a decomposed from:
\beq\label{eq:Integral_Dissection}
    I_N^{D} (Q) = \sum_{\cQ \in \operatorname{BS}(Q)} \sgn(\cS_{\cQ})  \frac{\sqrt{|\det \cQ|}}{\sqrt{|\det Q|}} \,  I_N^D (\cQ)\, ,
\eeq
which follows from the properties of \(\operatorname{BS}(Q)\) and was proven in ref.~\cite{Duhr:2025azh}.

Using eq.~\eqref{eq:Integral_Dissection}, the computation of \(I_N^D(Q)\) reduces to
the evaluation of dimensionally-regularised integrals \(I_N^D(\cQ)\) associated with orthoschemes \(\cQ \in \operatorname{BS}(Q)\).
We will refer to these Feynman integrals as \emph{orthoscheme Feynman integrals} or loosely as orthoschemes when there is no confusion. 

Note that, when \(Q\) is a generic Cayley matrix in Euclidean kinematics, then the orthoscheme Gram matrices of \(\operatorname{BS}(Q)\) fall into a special class of matrices, i.e., they fulfill
\beq\label{eq:ortho_form} \cQ_{i,j} = \cQ_{\max\{i,j\},\max\{i,j\}} \, , \eeq and the diagonal entries are ordered \(0<\cQ_{1,1} < \cQ_{2,2} < \ldots < \cQ_{N,N} < \infty\) 
(see ref.~\cite{Duhr:2025azh} for details).
We will refer to this class of matrices as the space of \emph{nested Gram matrices}, denoted
\beq\label{eq:cGnest_def}
\mathcal{G}_N^{\mathrm{nest}}
=
\Bigl\{ \cQ \in \mathbb R^{N\times N} \;\Big|\;
\cQ_{i,j} = \cQ_{\max\{i,j\},\max\{i,j\}}, \;
0<\cQ_{1,1}<\cdots<\cQ_{N,N}
\Bigr\}\, .
\eeq
In the remainder of this paper we therefore focus on integrals \(I_N^D(\cQ)\) with \(\cQ \in \mathcal{G}_N^{\mathrm{nest}}\), since the full integral \(I_N^D(Q)\) can be reconstructed from these building blocks via eq.~\eqref{eq:Integral_Dissection}. As we will show in the next section, the special structure of nested Gram matrices endows these integrals with additional properties that make them particularly amenable to a representation in terms of MPLs by linking them to limits of higher dimensional hyperbolic volumes. 
\section{Schläfli's equation and \texorpdfstring{\(\eps\)}{Epsilon}-recursions}\label{sec:schlaefli}
In this section we show how Schläfli's differential formula leads to recursive relations between $\eps$-expansion coefficients of one-loop $N$-point functions. The section is structured as follows. First we review Schläfli's formula, which gives a differential relation for volumes of hyperbolic simplices. 
Afterward, we apply this relation to obtain a formula expressing the first-order $\eps$ contribution of an $N$-point Feynman integral as the limit of an $(N+1)$-point integral.
We then extend this construction to arbitrary orders in $\eps$ and bridge the even and odd cases through a certain limiting procedure.

\subsection{Geometric Setup: Schl\"afli's differential equation}\label{sec:SchlaefliFormula}

\begin{figure}[htbp]
    \centering
    \includegraphics[width=0.5\linewidth]{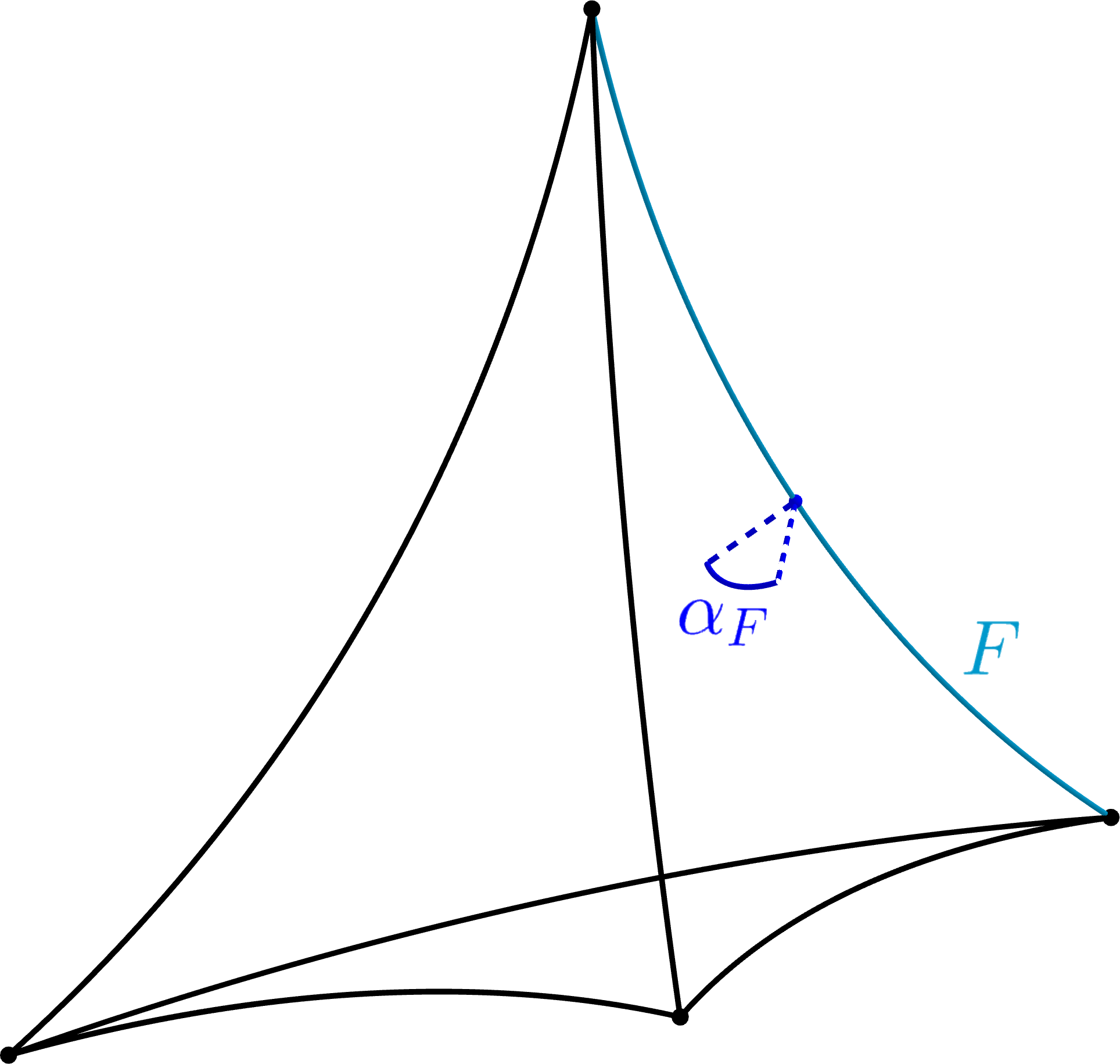}
    \caption{Illustration of a codimension-\(2\) face \(F\) within a three-dimensional simplex. The face \(F\) is the edge common to two facets, and \(\alpha_F\) is the dihedral angle between them.}
    \label{fig:definition_F}
\end{figure}

In contrast to Euclidean geometry, where volumes are determined by edge lengths, in hyperbolic geometry the volume of a simplex is governed by its dihedral angles. This relationship is captured by Schläfli's differential formula \cite{Schlafli1901,Kneser:1936simplex,kellerhals_volume_1989,kellerhals_schlaflis_1991}, i.e., for an $(N+1)$-dimensional hyperbolic simplex \(P\), the change in its volume is given by
\begin{equation}
    \mathrm{d} \! \Vol_{N+1} (P) = -\frac{1}{N} \sum_{F \subset P} \Vol_{N-1} (F)\, \mathrm{d} \alpha_F \, .
    \label{eq:Schläfli_DEQ}
\end{equation}
Here, the sum runs over all codimension-\(2\) faces \(F\) of the simplex, \(\Vol_{N-1}(F)\) denotes the volume of such a face, and \(\mathrm{d}\alpha_F\) is the differential of the corresponding dihedral angle (see figure~\ref{fig:definition_F}). 
Schläfli's formula thus provides a differential control of hyperbolic volumes in terms of their dihedral angles.

For a general hyperbolic simplex with Gram matrix \(Q\), the dihedral angle \(\alpha_{i,j}\) between the \(i\)-th and \(j\)-th facets is determined by the inverse Gram matrix via (see~ref.~\cite{Ren:2023tuj})
\begin{equation}
\left(\cos \alpha_{i,j} (Q) \right)^2
= \frac{(Q^{-1}_{i,j})^2}{Q^{-1}_{i,i} Q^{-1}_{j,j}}\, .
\label{eq:general_dihedral_angle}
\end{equation}
Although Schläfli's differential equation applies to arbitrary simplices and hence can be adapted to general Feynman integrals, its direct use becomes impractical due to the large number of independent dihedral angles. We, therefore, restrict our attention to the special class of orthoschemes, i.e., to nested Gram matrices.\footnote{This is justified because, by eq.~\eqref{eq:Integral_Dissection}, the general case can be reduced to this class.}
In this case, the number of non-trivial dihedral angles is significantly reduced, and their simple structure later allows Schläfli's differential equation to be integrated explicitly along a one-parameter deformation.

For orthoschemes, the only non-right dihedral angles are of the form \(\alpha_{i,i+1}\), which we denote by \(\alpha_i\) for brevity. For the special orthoschemes with nested Gram matrices, these angles take a particularly simple form. 
In the following, we need to consider an \((N+1)\)-point function for \(N\) even with an \((N+1)\times(N+1)\) Cayley matrix \(\cQ \in \mathcal{G}_{N+1}^\textnormal{nest}\). 
The associated Gram matrix is then given by \({\cQ}_\textnormal{ext}\), for which the non-trivial dihedral angles are given by
\begin{equation}
\label{eq:dihedral_angles_odd}
(\cos \alpha_i (\cQ_\textnormal{ext}) )^2 =
\begin{cases}
    \dfrac{\cQ_{2,2}-\cQ_{3,3}}{\cQ_{1,1}-\cQ_{3,3}}, & i=1, \\[2ex]
    \dfrac{(\cQ_{i-1,i-1}-\cQ_{i,i})(\cQ_{i+1,i+1}-\cQ_{i+2,i+2})}
    {(\cQ_{i-1,i-1}-\cQ_{i+1,i+1})(\cQ_{i,i}-\cQ_{i+2,i+2})}, & 1<i<N, \\[2ex]
    \dfrac{\cQ_{N-1,N-1}-\cQ_{N,N}}{\cQ_{N,N}-\cQ_{N+1,N+1}}, & i=N, \\[2ex]
    1 - \dfrac{\cQ_{N,N}}{\cQ_{N+1,N+1}}, & i=N+1.
\end{cases}
\end{equation}
For notational simplicity, we omit the subscript ``ext'' in geometric quantities (e.g., dihedral angles and volumes) whenever the intended object is clear from the dimension of the Gram matrix. In particular, when \( \cQ \) has odd dimension, it is understood that \( \cQ \) actually means \( \cQ_{\mathrm{ext}} \), such that we simply write \( \alpha_i(\cQ) \) instead of \( \alpha_i(\cQ_{\mathrm{ext}}) \).

A crucial feature of eq.~\eqref{eq:dihedral_angles_odd} is that, except for \(\alpha_{N+1}\), all dihedral angles depend only on differences of entries of \(\cQ\). Consequently, these angles are invariant under the shift deformation
\beq
\cQ \to \cQ_\eta = \cQ + \eta\,\mathbf{1}\mathbf{1}^\top\, ,
\eeq
i.e.,
\beq
\alpha_i(\cQ_\eta) = \alpha_i(\cQ), \qquad 1\le i \le N\, .
\eeq
This property will play an important role in the application of Schläfli's formula in the following sections.

\subsection{From Schläfli to the \texorpdfstring{\(\mathcal{O}(\eps)\)}{O(eps)} term}\label{sec:schlaefli_rec-1}
We now use the difference property of the dihedral angles in eq.~\eqref{eq:dihedral_angles_odd} to relate Schläfli’s differential formula to the \(\mathcal{O}(\eps)\) term of one-loop integrals. 
Let \(N\) be even and denote by \(\cQ\) the \(N\times N\) Cayley matrix of the corresponding orthoscheme. 
Furthermore, let \(P\) be the orthoscheme associated to an \((N+1)\)-point function with Cayley matrix \(\check{\cQ}\) such that
\(\check{\cQ}_{i,j}=\cQ_{i,j}\) for \(1 \le i,j \le N\) and \(\check{\cQ}_{i,N+1}=\cQ_{N+1,N+1}\), i.e.
\begin{equation}
\label{eq:Qext_def_regulated}
\check{\cQ} =
\begin{pmatrix}
\cQ & \cQ_{N+1,N+1}\,\mathbf{1}\\
\cQ_{N+1,N+1}\,\mathbf{1}^{\top} & \cQ_{N+1,N+1}
\end{pmatrix}\, .
\end{equation}

The orthoscheme \(P\) is constructed such that the dihedral angles are given by eq.~\eqref{eq:dihedral_angles_odd}. 
To make contact with the auxiliary-mass representation discussed in section~\ref{sec:auxiliary-mass}, we now introduce the one-parameter deformation of the Gram matrix obtained by shifting all entries according to
\(\check{\cQ}\to\check{\cQ}_{\eta} = \check{\cQ} + \eta \, \mathbf{1}\mathbf{1}^\top\).
Because of the difference structure discussed in section~\ref{sec:SchlaefliFormula}, the dihedral angles satisfy
\(\alpha_i(\check{\cQ}_{\eta})=\alpha_i(\check{\cQ})\) for \(i \le N\),
whereas the last angle is given by
\begin{equation}
\label{eq:last_dihedral}
(\cos\alpha_{N+1}(\check{\cQ}_{\eta}))^{2}
=\frac{\cQ_{N+1,N+1}-\cQ_{N,N}}{\cQ_{N+1,N+1}+\eta}\, .
\end{equation}

We now consider the variation of the simplex volume along this deformation. Since \(\alpha_{N+1}\) is the only dihedral angle that depends on \(\eta\), the sum in Schläfli's formula collapses to a single term. The codimension-two face conjugate to this angle is precisely the original orthoscheme associated with the shifted matrix \(\cQ_\eta\) (viewed as a face of the enlarged simplex), and therefore has volume \(\Vol_{N-1}(\cQ_\eta)\). 
Indeed, in the extended Gram matrix \((\check{\cQ}_{\eta})_{\mathrm{ext}}\), this face is obtained by deleting the final two rows and columns, leaving precisely \(\cQ_{\eta}\).
Thus Schläfli's differential formula reduces to
\begin{equation}
\label{eq:Schlafli_eta}
\mathrm{d}\! \Vol_{N+1}(\check{\cQ}_{\eta})
= -\frac{1}{N}\,\Vol_{N-1}(\cQ_{\eta})\,
\dv{\alpha_{N+1}}{\eta}\,\mathrm{d}\eta \, .
\end{equation}

The derivative of the dihedral angle is given by
\begin{equation}
\begin{split}
\dv{\alpha_{N+1}(\check{\cQ}_{\eta})}{\eta}
&= 
\dv{}{\eta}
\left[
\operatorname{arccos}\!\left(\sqrt{
\frac{\cQ_{N+1,N+1}-\cQ_{N+1,N+1}}{\cQ_{N,N}+\eta} }
\right)
\right] \\[1.5ex]
&= \frac{\sqrt{\cQ_{N+1,N+1}-\cQ_{N,N}}}
{2\sqrt{\cQ_{N,N}+\eta}\,(\cQ_{N+1,N+1}+\eta)} \, .
\end{split}
\end{equation}

To translate eq.~\eqref{eq:Schlafli_eta} into a relation between Feynman integrals, 
we use the relation between hyperbolic volumes and Feynman integrals in eq.~\eqref{eq:INN_Vol},
replacing \(\Vol_{N-1}(\cQ_{\eta})\) and \(\Vol_{N+1}(\check{\cQ}^{\eta}_{\textnormal{ext}})\) by
\(I_{N}^{N}(\cQ_{\eta})\) and \(I_{N+1}^{N+2}(\check{\cQ}_{\textnormal{ext}}^{\eta})\), respectively.
The corresponding Gram determinants are
\begin{align}
\det \left(\cQ_{\eta} \right)
&=(\cQ_{1,1}-\cQ_{2,2})\cdots(\cQ_{N-1,N-1}-\cQ_{N,N})(\cQ_{N,N}+\eta),\\[1.2ex]
\det\left( (\check{\cQ}_{\eta})_{\textnormal{ext}} \right) 
&=-\frac{1}{4}(\cQ_{1,1}-\cQ_{2,2})\cdots(\cQ_{N,N}-\cQ_{N+1,N+1})\, ,
\end{align}
so that
\begin{equation}
\frac{\det\left( (\check{\cQ}_{\eta})_{\textnormal{ext}} \right) }{\det \left(\cQ_{\eta} \right)}
= \frac{1}{4}\,
\frac{\cQ_{N+1,N+1}-\cQ_{N,N}}{\cQ_{N,N}+\eta}\, .
\end{equation}

After accounting for these determinant factors, Schläfli's formula, i.e., eq.~\eqref{eq:Schlafli_eta}, yields the integral relation
\begin{equation}
\label{eq:In+1toIn}
I_{N+1}^{N+2}(\check{\cQ})
= \frac{1}{2}\int_{0}^{\infty}\mathrm{d}\eta\,
\frac{I_{N}^{N}(\cQ_{\eta})}{\cQ_{N+1,N+1}+\eta}\, ,
\end{equation}
where \(\lim_{\eta \to \infty} I_{N+1}^{N+2}(\check{\cQ}_\eta)=0\) by power counting, allowing us to relate an \((N+1)\)-point function in \(N+2\) dimensions to an integral over a one-parameter family of \(N\)-point functions in \(N\) dimensions.

Equation~\eqref{eq:In+1toIn} is derived in the hyperbolic region. In the following we use its analytic continuation along the positive real \(q=\cQ_{N+1,N+1}\) axis to \(q\to0^+\).
For Euclidean kinematics the corresponding Feynman-parameter integrals remain well defined along this continuation, although their geometric interpretation changes at \(q=\cQ_{N,N}\). This geometric transition is illustrated explicitly for the bubble integral in section~\ref{app:geometric_transition}.

The connection to the \(\varepsilon\)-expansion follows from the observation that the right-hand side of eq.~\eqref{eq:In+1toIn}
provides an alternative representation of the auxiliary-mass integral discussed previously, in which the singular behaviour at \(\eta=0\) is regulated by the kinematic scale \(Q_{N+1,N+1}\) rather than by \(\eps\).
In particular, the coefficient of order \(\mathcal{O}(\eps)\) was given by (see eq.~\eqref{eq:eps_expansion_coeff}):
\begin{equation}\label{eq:first_eps_term_calI}
\mathcal{I}_{N}^{0}(\cQ)
= \int_{0}^{1}\frac{\mathrm{d}u}{u}\bigl[I_{N}^{N}(u)-I_{N}^{N}(0)\bigr]
+ \int_{0}^{1}\frac{\mathrm{d}u}{1-u}\,I_{N}^{N}(u)\, ,
\end{equation}
where we defined \(I_{N}^{N}(u) = I_N^N (\cQ_{u/(1-u)}) = I(u)\).
We can obtain the same object from eq.~\eqref{eq:In+1toIn} via a limit as we demonstrate now.

After introducing the variable transformation \(\eta = \tfrac{u}{1-u}\) and defining \( q = \cQ_{N+1,N+1} \) for brevity, we focus on the limit \(0 < q \ll 1\) and perform a series expansion of eq.~\eqref{eq:In+1toIn} with respect to \(q\): 
\beq\bsp\label{eq:In+1toIn_u}
2I_{N+1}^{N+2}(\check{\cQ}) &= \int_0^1 \mathrm{d}u \, \left(\frac{1}{(1- q ) u + q } + \frac{1}{1 - u} -\frac{q}{(1 - q ) u + q } \right) I_N^N (u) \\
&= \int_0^1 \mathrm{d}u \, \left( \frac{1}{ u + q } + \frac{1}{1 - u} \right)I_N^N(u)  + \mathcal{O}(q) \, .
\esp\eeq
As discussed in section~\ref{sec:auxiliary-mass}, the apparent pole from \(\tfrac{1}{1-u}\) is spurious and can be expanded under the integral sign without difficulty.
In contrast, the term \(\tfrac{1}{u+q}\) develops a singularity in the limit \(q \ll 1\), and therefore requires a more careful treatment.
In analogy with the distributional expansion of \(u^{-1-\eps}\) in section~\ref{sec:auxiliary-mass},
we can also derive a distributional expansion\footnote{
This follows by integrating against a test function \(\varphi(u)\):
\[
\int_0^1 \mathrm{d}u \, \frac{\varphi(u)}{u+q}
= \varphi(0)\int_0^1 \frac{du}{u+q}
+ \int_0^1 \mathrm{d}u\, \frac{\varphi(u)-\varphi(0)}{u+q} \, .
\]
The first term yields \(-\varphi(0)\log q + \mathcal{O}(q)\), while the second
tends to \(\int_0^1 \mathrm{d}u \, \frac{\varphi(u)-\varphi(0)}{u} + \mathrm{o}(1)\), which defines the plus-distribution.
}
of \(\tfrac{1}{u+q}\) around \(q=0\):
\beq\label{eq:dist_exp_1/u+q}
    \frac{1}{u+q} = - \log (q) \,  \delta (u) + \left[ \frac{1}{u} \right]_+ + \mathrm{o}(1)\, .
\eeq
Inserting eq.~\eqref{eq:dist_exp_1/u+q} into eq.~\eqref{eq:In+1toIn_u} and comparing with eq.~\eqref{eq:first_eps_term_calI}, we obtain:%
\begin{equation}
\label{eq:1eps=n+1gon}
2I_{N+1}^{N+2}(\check{\cQ})
= \mathcal{I}_{N}^{0}(\cQ)
- \log(q)\,I_{N}^{N}(0)
+ \mathrm{o}(1)\, .
\end{equation}
Rearranging terms and taking the limit, we find%
\begin{equation}
\label{eq:first_eps}
\mathcal{I}_{N}^{0}(\cQ)
= \lim_{\cQ_{N+1,N+1}\to0}
\left[
2I_{N+1}^{N+2}(\check{\cQ})
+ \log(\cQ_{N+1,N+1})\,I_{N}^{N}(0)
\right]\, .
\end{equation}
This identity expresses the \(\mathcal{O}(\eps)\) term of the Laurent expansion as a limit of a higher-point integral at \(\eps = 0\).
Since both \(I_{N+1}^{N+2}(\check{\cQ})\) and \(I_N^N(0)\) admit representations in terms of multiple polylogarithms, the same holds for \(\mathcal{I}_N^0(Q)\).
\subsection{All-orders \texorpdfstring{\(\eps\)}{Epsilon}-recursion from Schl\"afli}\label{sec:AllOrdersRecurrence}

Having established the geometric origin of the $\mathcal{O}(\varepsilon)$ coefficient in the Laurent expansion, we now generalise this result to construct a recursive relation for the coefficients at arbitrary order in $\varepsilon$.
The strategy is to introduce a generalised integral representation that makes the logarithmic kernels appearing in the Laurent expansion (see eq.~\eqref{eq:eps_expansion_coeff}) manifest, and then apply integration-by-parts to relate the coefficient at order \(k\) of an \(N\)-point function to the coefficient at order \(k-1\) of an \((N+1)\)-point function.
For the moment we assume \(k\geq1\). The case \(k=0\) was established separately in eq.~\eqref{eq:first_eps}.

Let \(q, \delta > 0\), and define
\begin{equation}\label{eq:IQqdelta}
\begin{split}
    \mathcal{I}_N^k (Q,q;\delta) \coloneqq \frac{(-1)^k}{k!} &\bigg(\int_0^1 \mathrm{d}u\, \left( \frac{1}{u+q} + \frac{1}{1-u} \right) \log \left( \frac{u+\delta}{1-u} \right)^k I_N^N (u) \\
    &\qquad - \int_0^1 \frac{\mathrm{d} u}{u+q} \log (u+\delta)^k I_N^N (0) \bigg) \, .
\end{split}
\end{equation}
We study the behaviour of this expression in the limit \(q, \delta \to 0\). While the expression in eq.~\eqref{eq:IQqdelta} is well-defined for \(\delta =0 \) we assume \(\delta > 0\) to regulate intermediate singularities arising during integration-by-parts. In the end \(\delta\) can be safely set to zero.
By construction, one recovers
\beq
\lim_{q \to 0} \mathcal{I}_N^k (Q,q;0) = \mathcal{I}_N^k (Q)\, .
\eeq

Our goal is to find a closed form for \(\mathcal{I}_N^k (Q,q;0)\) which we know how to evaluate. The expression in eq.~\eqref{eq:IQqdelta} consists of two parts. The first part is the main body of interest while the second term, involving \(I_N^N(0)\), contains all the singularities in \(q\) to cancel all divergences of the first. We start by treating the latter. In the \(\delta \to 0\) limit, this integral can be evaluated in terms of polylogarithms:
\begin{equation}\label{eq:divergent_part_integral}
    \begin{split}
    I_2 &\coloneqq \lim_{\delta \to 0} \int_0^1 \frac{\mathrm{d}u}{u+q} \log(u+\delta)^k\, I_N^N(0) = I_N^N(0) \int_0^1 \frac{\mathrm{d}u}{u+q} \log(u)^k\\[1.25ex]
    &= I_N^N(0) ~ k!\, G(-q,\underbrace{0,\ldots,0}_{k\textnormal{ times}};1) = (-1)^{k+1} I_N^N(0) ~ k!\, \operatorname{Li}_{k+1} \left(- \frac{1}{q}\right)\, . %
    \end{split}
\end{equation}

The polylogarithm \(\operatorname{Li}_{k+1} \left(- \tfrac{1}{q}\right)\) in eq.~\eqref{eq:divergent_part_integral} is divergent at \(q=0\), and, for small \(q>0\), has the known asymptotic expansion:
\begin{equation}\label{eq:divergencies}
    \begin{split}
         (-1)^{k+1} \operatorname{Li}_{k+1} \left(- \frac{1}{q}\right) = \sum_{i=0}^{\big\lfloor \tfrac{k+1}{2} \big\rfloor} 2(2^{1-2i}-1) \zeta(2i) \frac{\log(q)^{k+1-2i}}{(k+1-2i)!}  \, ,
    \end{split}
\end{equation}
where we use the standard convention \(\zeta(0) = -\tfrac{1}{2}\).

We now show that the remaining integral reproduces the integral representation of \(\mathcal{I}_{N+1,N+2}^{k-1}(\check{\cQ})\) as \(\delta \to 0^+\), which is the origin of the recursion. In particular, we simplify the remaining integral through integration-by-parts.
The key insight is to recognize that the primitive appearing in the integration-by-parts step is directly related to the \((N+1)\)-point function integral \(I_{N+1}^{N+2} (u)\) via eq.~\eqref{eq:In+1toIn}. We obtain:
\begin{equation}
    \begin{split}
        I_1 &\coloneqq \int_0^1 \mathrm{d}u\, \left( \frac{1}{u+q} + \frac{1}{1-u} \right) \log \left( \frac{u+\delta}{1-u} \right)^k I_N^N (u)\\
        &= \log \left( \frac{u+\delta}{1-u} \right)^k \int_1^u \mathrm{d}v \left( \frac{1}{v+q} + \frac{1}{1-v} \right)I_N^N (v) \bigg|_{u=0}^1 \\
        &\quad - k \int_0^1 \mathrm{d} u\, \left( \frac{1}{u+\delta} + \frac{1}{1-u} \right)\log \left( \frac{u+\delta}{1-u} \right)^{k-1} \int_1^u \mathrm{d}v \left( \frac{1}{v+q} + \frac{1}{1-v} \right)I_N^N (v)\, .
    \end{split}
\end{equation}
In the derivation of eq.~\eqref{eq:In+1toIn}, we could have specified a non-zero lower bound \(\eta'\) generalising to:
\begin{equation}
\label{eq:In+1toIn_gen}
I_{N+1}^{N+2}(\check{\cQ}_{\eta'})
= \frac{1}{2}\int_{\eta'}^{\infty}\mathrm{d}\eta\,
\frac{I_{N}^{N}(\cQ_{\eta})}{\cQ_{N+1,N+1}+\eta}\, .
\end{equation}
For \(\eta'=\tfrac{u}{1-u}\) we can then similarly to eq.~\eqref{eq:In+1toIn_u} derive an expression in the variables \(\eta = \tfrac{v}{1-v}\) to obtain:
\beq\label{eq:In+1toIn_gen_u}
\int^1_u \mathrm{d}v \left( \frac{1}{v+q} + \frac{1}{1-v} \right)I_N^N (v) = 2 I_{N+1}^{N+2}(u) \, ,
\eeq
to leading order in \(q\).
We then use this relation to rewrite the primitive as an \((N+1)\)-point function in the integration-by-parts step, which yields
\beq\label{eq:I1_inter}
I_1 = 2 \log (\delta )^k \,I_{N+1}^{N+2} (0) + 2k \int_0^1 \mathrm{d} u\, \left( \frac{1}{u+\delta} + \frac{1}{1-u} \right)\log \left( \frac{u+\delta}{1-u} \right)^{k-1} I^{N+2}_{N+1} (u) \, .
\eeq
To proceed, we expand the logarithm in the integrand binomially,
\begin{equation}
\label{eq:log_binomial}
    \log\!\left(\frac{u+\delta}{1-u}\right)^{k-1} = \sum_{j=0}^{k-1} \binom{k-1}{j} \log(u+\delta)^{j}\,(-\log(1-u))^{k-1-j}\, ,
\end{equation}
and apply the generalised distributional expansion (compare with eq.~\eqref{eq:dist_exp_1/u+q})
\begin{equation}
\label{eq:dist_exp_log_power}
    \frac{\log(u+\delta)^n}{u+\delta} = -\frac{1}{n+1}\log(\delta)^{n+1}\,\delta(u) + \left[\frac{\log(u)^n}{u}\right]_+ \, .
\end{equation}
Only the $j=k-1$ term in eq.~\eqref{eq:log_binomial} contributes to the $\delta(u)$ piece, since the factor $\log(1-u)^{k-1-j}$ vanishes at $u=0$ for $j < k-1$. The $\delta(u)$ contribution to the bulk integral in eq.~\eqref{eq:I1_inter} therefore evaluates to
\( -2\log(\delta)^{k}\,I_{N+1}^{N+2}(0)\),
which cancels the logarithmic boundary term $2\log(\delta)^{k}\,I_{N+1}^{N+2}(0)$ in eq.~\eqref{eq:I1_inter} exactly; hence, $I_1$ is finite as $\delta \to 0^+$. The remaining $[\log(u)^n/u]_+$ contributions combine with the $1/(1-u)$ piece of the integrand to yield:
\begin{equation}
    \begin{split}
        I_1 &= 2k \int_0^1 \frac{\mathrm{d}u}{u} \left( \log \left( \frac{u }{1-u} \right)^{k-1} I^{N+2}_{N+1} (u) - \log(u)^{k-1} I^{N+2}_{N+1} (0) \right)\\
        &\quad + 2k \int_0^1 \frac{\mathrm{d} u}{1-u} \log\left( \frac{u}{1-u} \right)^{k-1} I^{N+2}_{N+1} (u)  + \mathcal{O}(\delta) \\[1.25ex]
        &= 2k!\, (-1)^{k-1} \, \mathcal{I}_{N+1,N+2}^{k-1} (\check{Q}) + \mathcal{O}(\delta)\, ,
    \end{split}
\end{equation}
where we used eq.~\eqref{eq:eps_expansion_coeff} in the last step to connect the integrals to the \(\eps\) expansion coefficients.
We see that \(I_1\) reproduces precisely the \((k-1)\)-order of an \((N+1)\) point function, while \(I_2\) 
contains precisely the terms required to cancel the singular behaviour of \(I_1\) in the limit \(q \to 0^+\).

Again combining \(I_1\) and \(I_2\) as in eq.~\eqref{eq:IQqdelta}, while setting \(\delta = 0\) and then taking the limit \(q \to 0\) yields the final result:
\begin{equation}\label{eq:recursion}
    \mathcal{I}_N^k (Q)= \lim_{q \to 0} \mathcal{I}_N^k (Q,q;0) = \lim_{q \to 0} \left(- 2\, \mathcal{I}_{N+1,N+2}^{k-1} (\check{Q})+ I_N^N (0) \Li_{k+1} \left(-\frac{1}{q} \right) \right)\, .
\end{equation}
The polylogarithmic term \(\Li_{k+1} \left(-\tfrac{1}{q} \right)\) encodes precisely the logarithmic divergences in \(q\) of \(\mathcal{I}_{N+1,N+2}^{k-1} (\check{Q})\) such that eq.~\eqref{eq:recursion} is well-defined.
To summarize, this relation links the 
\(\mathcal{O}(\eps^{k+1})\) term
of an \(N\)-point function (for \(N\) even) to the \(\mathcal{O}(\eps^k)\) term of an \((N+1)\)-point function.\footnote{Recall that the order in the \(\eps\)-expansion of the Feynman integrals is shifted by \(+1\) with respect to \(k\) in the quantities \(\mathcal{I}_{N,d}^k\).}
The derivation above assumes \(k\geq1\). Using \(\mathcal I_{N,d}^{-1}=-I_N^d\), eq.~\eqref{eq:recursion} also applies for \(k=0\), where it reduces to the first-order relation in eq.~\eqref{eq:first_eps}.
It therefore generalises the work of section~\ref{sec:schlaefli_rec-1}, i.e., eq.~\eqref{eq:first_eps}, which is recovered for \(k=0\).

To close the recursion the only loose end is how to obtain \(\mathcal{I}_{N-1,N}^{k} (\cQ)\) since our recursion only gives results for Feynman integrals with an even number of external legs.
This can be achieved by a `large-mass limit', i.e., the following relation holds:
\begin{equation}\label{eq:eventoodd}
    \mathcal{I}_{N-1,N}^k (\cQ) = \lim_{\cQ_{N,N}\, \to \, \infty} 2 Q_{N,N} \, \mathcal{I}_{N}^k (\check{\cQ})\, , \qquad N\geq4\, .
\end{equation}
Here \(\cQ\) is a \((N-1) \times (N-1)\) nested Gram matrix and the additional factor \(2 Q_{N,N}\) is an adjustment accounting for slightly different determinants. We postpone the proof to appendix~\ref{app:evenodd}.

Together, the Schläfli-recursion~\eqref{eq:recursion} and the even-to-odd-limit~\eqref{eq:eventoodd} form a complete and powerful algorithmic framework. Because of these two procedures, we can recursively compute any \(\eps\)-order term for any one-loop integral, from the \(\eps = 0\) expressions (see ref.~\cite{Ren:2023tuj}) with a sufficient number of legs.

\begin{figure}[htbp!]
    \centering
    \begin{tikzpicture}[
        dot/.style={circle, fill=black, inner sep=1.5pt},
        dotgray/.style={circle, fill=gray, inner sep=1.5pt},
        main arrow/.style={->, thick, shorten >=2pt, shorten <=2pt},
        dashed arrow/.style={->, dashed, color=gray, shorten >=2pt, shorten <=2pt},
        node distance=1.2cm and 1.5cm
    ]

    \foreach \i [evaluate=\i as \ni using int(8-\i)] in {0,...,6} {
        \node[left=0.5cm] at (0,-\i) {$N=\ni$};
    }

    \foreach \i in {0,...,6} {
        \node[above=0cm] at (\i*1.5,0.5) {$\varepsilon^{\i}$};
    }

    \foreach \i in {0,...,6} {
        \foreach \j in {0,...,6} {
            \node[dotgray] (n\i\j) at (\j*1.5, -\i) {};
        }
    }

    \foreach \i in {0,...,6} {
        \pgfmathtruncatemacro{\ip}{\i+1}
        \foreach \j in {0,...,5} {
            \pgfmathtruncatemacro{\n}{7 - \i} %
            \pgfmathtruncatemacro{\np}{7 - \ip}
            \ifnum\n>1
                \ifodd\n
                    \draw[dashed arrow] (n\i\j) -- (n\ip\j);
                \else
                    \pgfmathtruncatemacro{\jp}{\j+1}
                    \draw[dashed arrow] (n\i\j) -- (n\ip\jp);
                \fi
            \fi
        }
    }
    
    \draw[dashed arrow] (n46) -- (n56);
    \draw[dashed arrow] (n26) -- (n36);
    \draw[dashed arrow] (n06) -- (n16);
    
    \foreach \i in {0,...,3} {
        \pgfmathtruncatemacro{\ip}{\i+1}
        \pgfmathtruncatemacro{\ipp}{\i+2}
        \pgfmathtruncatemacro{\ippp}{4 - \i}
        \foreach \j in {0,...,\ippp} {
            \pgfmathtruncatemacro{\n}{7 - \i} %
            \pgfmathtruncatemacro{\np}{7 - \ip}
            \pgfmathtruncatemacro{\jp}{\j+1}
            \pgfmathtruncatemacro{\ij}{\i+\j}
            \pgfmathtruncatemacro{\ijp}{\i+\j+1}
            \pgfmathtruncatemacro{\ijpp}{\i+\j+2}
            \pgfmathtruncatemacro{\k}{\j / 2}
            \pgfmathtruncatemacro{\kp}{\k+1}
            \ifnum\n>1
                \ifodd\ijp
                    \node[dot] (n\ij\k) at (\k*1.5, -\ij) {};
                    \node[dot] (n\ijp\k) at (\k*1.5, -\ijp) {};
                    \node[dot] (n\ijpp\kp) at (\kp*1.5, -\ijpp) {};
                    \draw[main arrow] (n\ij\k) -- (n\ijp\k);
                    \draw[main arrow] (n\ijp\k) -- (n\ijpp\kp);
                \else
                    \node[dot] (n\ij\k) at (\k*1.5, -\ij) {};
                    \node[dot] (n\ijp\kp) at (\kp*1.5, -\ijp) {};
                    \node[dot] (n\ijpp\kp) at (\kp*1.5, -\ijpp) {};
                    \draw[main arrow] (n\ij\k) -- (n\ijp\kp);
                    \draw[main arrow] (n\ijp\kp) -- (n\ijpp\kp);
                \fi
            \fi
        }
    }
    
    \node[dot] (n60) at (0, -6) {};

    \node[draw=red, thick, fit=(n00)(n60), inner sep=5pt] {};
    \node[draw=green, dashed, thick, fit=(n31)(n61)(n36)(n66), inner sep=8pt] {};
    
    \end{tikzpicture}
    \caption{Recursive computation scheme for \(\eps\)-terms in the Laurent expansion. 
    The \(\eps\)-order refers to the expansion order of the Feynman integral, not to the index of the object $\mathcal{I}^k_{N,d}(\cQ)$. Diagonal arrows represent the recursion \eqref{eq:recursion}, while vertical arrows correspond to the even-to-odd conversion \eqref{eq:eventoodd}. 
    Boxed red entries are known from \cite{Ren:2023tuj} (valid for all $N$), and green dashed-boxed entries from \cite{Duhr:2025azh} (valid to all orders in \(\eps\)). 
    Bold black entries are explicitly computable given integer-dimensional Feynman integrals up to the octagon ($N=8$).
    Gray dashed arrows indicate entries requiring higher $N$-point functions ($N\ge 9$) within this schematic illustration.   
    }
    \label{fig:tikz_flowchart}
\end{figure}
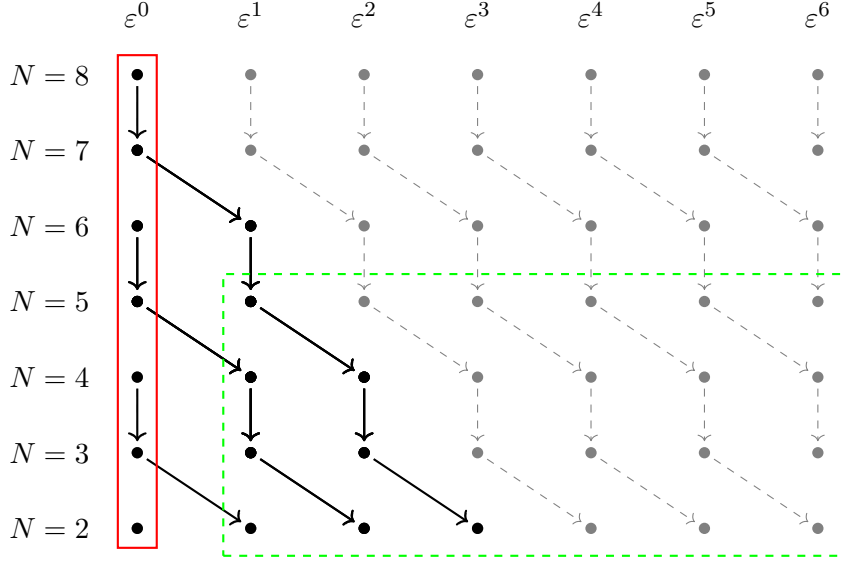

The diagram in figure~\ref{fig:tikz_flowchart} schematically depicts this twofold recursive procedure.
It shows how to obtain all \(\eps\)-terms in the Laurent expansion from the \(\eps=0\) contribution via sequential applications of the limits \eqref{eq:recursion} (diagonal arrows) and \eqref{eq:eventoodd} (vertical arrows). 

\section{Consequences of the recursion}
\label{sec:consequences}
Having established our main result in eq.~\eqref{eq:recursion}, we now discuss some structural consequences in this section before we apply the method to some examples in section~\ref{app:examples} and the hexagon in section~\ref{sec:hexagon}.

\subsection{One-loop integrals and MPLs}\label{sec:closure}
The first important takeaway from the recursion derived above is that all Laurent coefficients in the \(\eps\)-expansion of one-loop Feynman integrals can be expressed in terms of multiple polylogarithms (MPLs). In particular, it is well-known that any such integral can be reduced to a (combination of) scalar Feynman integrals \(I_N^{d-2\eps} (Q)\) with \(d = N\) for \(N\) even and \(d = N+1\) for \(N\) odd. This is a consequence of combining tensor reduction, integration-by-parts, and dimensional recurrence relations. 

The integrals \(I_N^{d-2\eps} (Q)\) can then be systematically treated through the framework developed in this paper. Namely, the Laurent coefficients \(\mathcal{I}_{N,d}^k (Q)\) which are connected to \(I_N^{d-2\eps} (Q)\) via
\beq
    I_N^{d-2\eps} (Q) = \frac{e^{\gamma_\textnormal{E} \eps}}{\Gamma (-\eps)} \sum_{k=-1}^\infty \mathcal{I}_{N,d}^k (Q) \eps^k\, ,
\eeq
are all expressible in terms of MPLs and are of uniform weight
\beq\label{eq:weight_formula}
    \operatorname{w}\!\left(\mathcal{I}_{N,d}^{k} ({Q})\right) = \frac{d}{2} + 1 + k\, .
\eeq
\noindent
This follows from the following observations:
\begin{enumerate}
\item \emph{Feynman integrals as sums of orthoschemes.}
Any scalar one-loop Feynman integral \(I_N^{d-2\eps}(Q)\) can be dissected into a sum of orthoscheme Feynman integrals \(I_N^{d-2\eps}(\cQ)\). 
Thus, determining the Laurent coefficients \(\mathcal{I}_{N,d}^k(\cQ)\) suffices to compute \(\mathcal{I}_{N,d}^k(Q)\).

\item \emph{Integer-dimensional integrals.}
For integer space-time dimension \(\eps = 0\), one-loop integrals \(I_N^d(\cQ)\) admit a geometric interpretation in terms of volumes of hyperbolic simplices. In Euclidean kinematics, these volumes can be expressed in terms of MPLs~\cite{Rudenko2020Orthoschemes,Ren:2023tuj} of weight \(\operatorname{w}\!\left( I_N^d(\cQ) \right) = \tfrac{d}{2}\). Hence, the integrals \(I_N^d(\cQ)\) are MPLs. Analytic continuation to other kinematic regimes is possible (see for example ref.~\cite{Duhr:2025azh}).

\item \emph{Recursive structure.}
The recursion relation in eq.~\eqref{eq:recursion} expresses the Laurent coefficients \(\mathcal{I}_N^k(\cQ)\) (\(N\) even) in terms of the integral \(\mathcal{I}_{N+1}^{k-1}(\check{\cQ})\) and subtraction terms. Each recursive step introduces one additional integration and increases the transcendental weight by one. Furthermore, the Laurent coefficient of Feynman integrals with an odd number of external legs can be derived from Feynman integrals with an even number of legs (see eq.~\eqref{eq:eventoodd}) through a limit preserving weight. 

\item \emph{Iterated result.}
Iterating this relation reduces any Laurent coefficient \(\mathcal{I}_{N,d}^k(\cQ)\) to contributions involving only integer-dimensional integrals, i.e., integrals at \(\eps=0\) with more external legs. 
In particular, \(\mathcal{I}_{N,d}^k(\cQ)\) can be related to \(I_{d+1+2k}^{d+2+2k}(\check{\cQ})\) which is of weight \(\operatorname{w}\!\left(I_{d+1+2k}^{d+2+2k}(\check{\cQ})\right) = \tfrac{d}{2} + 1 + k\).

\item \emph{Stability under limits and analytic continuation.}
The recursion expresses \(\mathcal{I}_{N,d}^k(\cQ)\) as limits of functions obtained from higher-point integrals at \(\eps=0\). 
In the limits appearing in the recursion, the logarithmic singular terms are explicitly isolated by the subtraction term in eq.~\eqref{eq:recursion}; the remaining finite part is a regularised limit of MPLs and remains within the same function class~\cite{Panzer:2014caa}. The weight is likewise preserved, so that \(\operatorname{w}\!\left(\mathcal{I}_{N,d}^k(\cQ)\right)=\operatorname{w}\!\left(I_{d+1+2k}^{d+2+2k}(\check{\cQ})\right)=\tfrac{d}{2}+1+k\). Analytic continuation changes the branches of these multivalued MPLs but does not enlarge the function class~\cite{Zhao:2003mpl}.

\end{enumerate}
\noindent
It follows that all Laurent coefficients of one-loop Feynman integrals are expressible as MPLs of uniform transcendental weight.
Configurations obtainable as finite or regularised limits of these expressions, including the massless limits considered here, remain in the same function class.

\subsection{Differential Equations}\label{app:deq}
It is instructive to compare our result with the differential equation derived in ref.~\cite[eq.~(9.9)]{Abreu:2017mtm} from the conjectured diagrammatic coaction. We show that, after specialization to orthoscheme kinematics, our expression reproduces the corresponding \(\mathcal{O}(\eps)\) differential equation, providing a non-trivial consistency check of the geometric construction. We start by considering even multiplicity and comment on the odd case afterwards.
Hence, unless stated otherwise, we assume that \(N\) is even.

\paragraph{Review of the known differential equations:}
Let us begin by recalling the known structure. In ref.~\cite{Abreu:2017mtm} a differential equation which includes the \(\mathcal{O}(\eps)\) term of scalar one-loop integrals with \(N\) external legs was provided. 
Using the notation of ref.~\cite{Abreu:2017mtm}, let \(G\) be the scalar one-loop Feynman graph with \(N\) external legs, then
we denote by \(J_G\) its associated Feynman integral in \(D=d-2\eps\) dimensions and leading singularity removed, i.e.,
\begin{equation}
    J_G = \frac{I_G}{j_N}, \quad \textnormal{with}\quad
    j_N = \begin{cases}
        2^{1- \frac{N}{2}} \,\iu^{\frac{N}{2}} \ Y_{[N]}^{-\frac{1}{2}} & N  \textnormal{ even,} \\[1.25ex]
        2^{\frac{1-N}{2}} \,\iu^{\frac{N-1}{2}} \ \Gram_{[N]}^{-\frac{1}{2}} & N \textnormal{ odd,}
    \end{cases}
\end{equation}
where \(I_G = I_N^D (Q)\), \(Y_{[N]} = Y_G = \det Q\), and \(\Gram_{[N]} = \Gram_G = -4 \det Q_\textnormal{ext} \). We assume that the kinematic information, i.e., \(Q\) is implicitly encoded in the (labelled) graph \(G\). %
The differential equation with this notation reads:
\begin{equation}
\label{eq:abreu_diff_eq}
    \mathrm{d} J_G = \eps J_G \, \mathrm{d} \mathcal{C}_{E_G}^{(1)} J_G
    + \sum_{\substack{X \subset E_G \\ n_X = n_G - 1}} \eps J_{G_X} \left( \mathrm{d} \mathcal{C}_X^{(1)} J_G + \tfrac{1}{2} \mathrm{d} \mathcal{C}_{E_G}^{(1)} J_G \right)
    + \sum_{\substack{X \subset E_G \\ n_X = n_G - 2}} J_{G_X} \, \mathrm{d} \mathcal{C}_X^{(0)} J_G\, ,
\end{equation}
where \(G_X\) is the graph contracted to the set of edges \(X \subset E_G\). 
The remaining terms, which we refer to as the \(\mathcal{C}\)-terms, are given by:
\begin{equation}
\label{eq:abreu_c_terms}
\begin{aligned}
    \mathcal{C}_{E_G}^{(1)} J_G &= \log \left( \frac{\operatorname{Gram}_{E_G}}{4 Y_{E_G}} \right), \\
    \mathcal{C}_{E_G \setminus e}^{(1)} J_G &= - \log \left( 1 + \sqrt{1 - \frac{\operatorname{Gram}_{E_G} Y_{E_G \setminus e}}{Y_{E_G} \operatorname{Gram}_{E_G \setminus e}}} \right)
    - \frac{1}{2} \log \left( \frac{\operatorname{Gram}_{E_G \setminus e}}{4 Y_{E_G \setminus e}} \right), \\
    \mathcal{C}_{E_G \setminus\{e, f\}}^{(0)} J_G &= \frac{1}{2} \log \left(
    \frac{
        \sqrt{Y_{E_G \setminus e} Y_{E_G \setminus f} - Y_{E_G} Y_{E_G \setminus\{e, f\}}}
        + \sqrt{-Y_{E_G} Y_{E_G \setminus\{e, f\}}}
    }{
        \sqrt{Y_{E_G \setminus e} Y_{E_G \setminus f} - Y_{E_G} Y_{E_G \setminus\{e, f\}}}
        - \sqrt{-Y_{E_G} Y_{E_G \setminus\{e, f\}}}
    } \right),
\end{aligned}
\end{equation}
with the subscript denoting restriction to the corresponding subset of edges, e.g., \(Y_X = \det Q_X\) where \(Q_X = Q_{X \times X} = (Q_{i,j})_{i,j \in X}\).

\paragraph{Evaluation for the Special Kinematics \texorpdfstring{\(\cQ\)}{Q}\,:}
We now specialize these general formulas to our orthoscheme kinematics, where the Gram matrix takes the orthoscheme form \(\cQ \in \mathcal{G}_N^{\mathrm{nest}}\) (see eq.~\eqref{eq:cGnest_def}). The differentials of the \(\mathcal{C}\)-terms simplify considerably:
\begin{align}
    \mathrm{d} \mathcal{C}_{E_G}^{(1)} J_G &= - \mathrm{d} \log \cQ_{N,N}, \\
    \mathrm{d} \mathcal{C}_{E_G \setminus e}^{(1)} J_G &= 
    \begin{cases}
        \tfrac{1}{2} \mathrm{d} \log \cQ_{N,N}, & e \neq N, \\
        \tfrac{1}{2} \mathrm{d} \log \cQ_{N-1,N-1} - \mathrm{d} \log \left( 1 + \sqrt{1 - \tfrac{\cQ_{N-1,N-1}}{\cQ_{N,N}}} \right), & e = N,
    \end{cases} \\
    \mathrm{d} \mathcal{C}_{E_G \setminus\{e, f\}}^{(0)} J_G &= 2 \iu \, \mathrm{d} \theta_{E_G \setminus\{e, f\}}\, ,
\end{align}
where \(\theta_{X}\) is the dihedral angle at the codimension two simplex associated to \(G_X\) in \(G\).

Observe that for \(\varepsilon=0\), one recovers Schläfli's differential equation as expected. We find that the combination appearing in the \(\varepsilon\)-term satisfies:
\begin{equation}
    \mathrm{d} \mathcal{C}_{E_G \setminus e}^{(1)} J_G + \frac{1}{2} \mathrm{d} \mathcal{C}_{E_G}^{(1)} J_G = 
    \begin{cases}
        0, & e \neq N, \\
        \frac{1}{2} \mathrm{d} \log \left( \frac{1 - x}{1 + x} \right), & e = N,
    \end{cases}
\end{equation}
where \(x = \sqrt{\tfrac{\cQ_{N,N} - \cQ_{N-1,N-1}}{\cQ_{N,N}}}\). 

\paragraph{Laurent expansion at \(\eps=0\):}
Expanding \(J_G = J_G^{(0)} + \varepsilon J_G^{(1)} + \ldots\), the first-order term in \(\eps\) satisfies:
\begin{equation}
\label{eq:our_diff_eq}
    \mathrm{d} J_G^{(1)} = - J_G^{(0)} \mathrm{d} \log \cQ_{N,N}
    + \tfrac{1}{2} J^{(0)}_{G_{E_G \setminus \{N\}}} \mathrm{d} \log \left( \frac{1 - x}{1 + x} \right)
    + \sum_{e=1}^{N-1} J^{(1)}_{G_{E_G \setminus \{e, e+1\}}} 2 \iu \, \mathrm{d} \theta_{E_G \setminus\{e, e+1\}}\, .
\end{equation}
This form is particularly powerful as it allows the symbol of the function \(J_G^{(1)}\) to be read off easily, which encodes its essential analytic properties.

\paragraph{Verification using the limiting formula:}
Finally, we demonstrate that the expression derived from our geometric limit, eq. \eqref{eq:first_eps}, satisfies the exact same differential equation. We write \(q=\cQ_{N+1,N+1}\) for brevity and define the quantity:
\begin{equation}\label{eq:calJG}
    \mathcal{J}_G^0(q) \coloneqq - J_{\check{G}} - \log(q) \, J_G \, .
\end{equation}
Here \(\check{G}\) is the graph associated to \(\check{\cQ}\) such that \(J_{\check{G}}\) depends on the parameter \(q\). By construction eq.~\eqref{eq:calJG}, gives \(\lim_{q\to 0} \mathcal{J}_G^0 (q) = J^{(1)}_G \) according to eq.~\eqref{eq:first_eps}.
Applying Schläfli's differential equation to eq.~\eqref{eq:calJG}, we obtain:
\begin{equation}\label{eq:deq-dJ0N}
\begin{split}
    \mathrm{d} \mathcal{J}_G^0 (q) &= \sum_{e=1}^{N-1}  2\iu \left(- J_{\check{G}_{E_{\check{G}} \setminus \{e,e+1\}}} \mathrm{d} \check{\theta}_{E_{\check{G}} \setminus\{e, e+1\}} - \log (q) \, J_{G_{E_G \setminus \{e,e+1\}}} \mathrm{d} \theta_{E_G \setminus\{e, e+1\}}\right)\\
    &\quad + 2\iu\, \left(  J_{\check{G}_{E_{\check{G}} \setminus \{N,N+1\}}} \mathrm{d} \check{\theta}_{N} - J_G \, \mathrm{d} \check{\theta}_{N+1} \right) - J_G\, \mathrm{d} \log (q)\, .
\end{split}
\end{equation}
Here \(\check{\theta}_e = \check{\theta}_{E_{\check{G}} \setminus\{e, e+1\}}\) are the dihedral angles of \(\check{\cQ}\).
In the limit \(q \to 0^+\), we find:
\begin{align}
    \check{\theta}_{E_{\check{G}} \setminus\{e, e+1\}} &\to \theta_{E_G \setminus\{e, e+1\}}, \\
    \check{\theta}_{N} = \theta_{E_{\check{G}} \setminus\{N, N+1\}} &\to \frac{\iu}{2} \log \left( \frac{1 - x}{1 + x} \right), \\
    \mathrm{d} \check{\theta}_{N+1} = \theta_{E_{\check{G}}\setminus\{N+1, N+2\}} &\to \frac{\iu}{2} \mathrm{d} \log (q)- \frac{\iu}{2} \mathrm{d} \log \cQ_{N,N}\, .
\end{align}
The first line in eq.~\eqref{eq:deq-dJ0N} can be rewritten, using eq.~\eqref{eq:recursion}, as:
\beq\bsp
    &\quad\lim_{q \to 0^+}  \left( - J_{\check{G}_{E_{\check{G}} \setminus \{e,e+1\}}} \mathrm{d} \check{\theta}_{E_{\check{G}} \setminus\{e, e+1\}} - \log (q) \, J_{G_{E_G \setminus \{e,e+1\}}} \mathrm{d} \theta_{E_G \setminus\{e, e+1\}} \right) \\
    &= \lim_{q \to 0^+} \mathcal{J}_{G_{E_G \setminus \{e,e+1\}}}^0 (q) \, \mathrm{d} \theta_{E_G \setminus\{e, e+1\}}   = J^{(1)}_{G_{E_G \setminus \{e,e+1\}}}\mathrm{d} \theta_{E_G \setminus\{e, e+1\}}  \, .
\esp\eeq
Furthermore, the terms in the second line simplify such that they reproduce \(- J_G^{(0)} \mathrm{d} \log \cQ_{N,N}\) and \( \tfrac{1}{2} J^{(0)}_{G_{E_G \setminus \{N\}}} \mathrm{d} \log ( \tfrac{1 - x}{1 + x} )\). In total,
we recover the differential equation in eq.~\eqref{eq:our_diff_eq}, confirming the consistency of our limiting approach with the differential equations of~\cite{Abreu:2017mtm}.

\paragraph{Even to odd limit:}
The case of odd multiplicity follows from the even case.
In particular, when \(N \) is even the \(N-1\) case can be obtained by considering:
\begin{equation}
    \frac{1}{2} \log \left( \frac{1-x}{1+x} \right) \to \frac{1}{2} \log \left( \frac{\cQ_{N-1,N-1}}{4 \cQ_{N,N}} \right) \textnormal{ as } \cQ_{N,N} \to \infty \, ,
\end{equation}
where the divergent part cancels with the contribution from \(\mathcal{C}_{E_G}^{(1)} J_G\) and the remainder is \(\log \cQ_{N-1,N-1}\). In particular, from eq.~\eqref{eq:our_diff_eq}, we obtain:
\begin{equation}
    \mathrm{d} J_{G'}^{(1)} = - J_{G'}^{(0)} \mathrm{d} \log \cQ_{N-1,N-1}
    + \sum_{e=1}^{N-1} J^{(1)}_{G'_{E_{G'} \setminus \{e, e+1\}}} 2\iu\, \mathrm{d} \theta_{E_{G'} \setminus\{e, e+1\}}\, ,
\end{equation}
where \(G'\) is the graph of the \(N-1\)-point function.
Comparing with the corresponding \(\mathcal{O}(\eps)\) result in ref.~\cite{Abreu:2017mtm} we find complete agreement. 
\section{Examples}\label{app:examples}
We illustrate our method for the examples of bubble, triangle, and box integrals. The hexagon is described in its own section \ref{sec:hexagon}. Since the triangle diagram is derived from the box via eq.~\eqref{eq:eventoodd} we treat the box prior to the triangle. The finite bubble diagram is studied as an instructive example up to \(\mathcal{O}(\eps^2)\) using our method. Furthermore, we present a geometric interpretation for its \(\mathcal{O}(\eps)\) term as volumes in Anti-de-Sitter space. This geometric interpretation can be extended to higher dimensional cases and deepens our understanding of dimensional regularised one-loop Feynman integrals.

\subsection{The bubble up to \(\mathcal{O}(\eps^2)\)}
We compute the \(\mathcal{O}(\eps)\) and \(\mathcal{O}(\eps^2)\) term of the two-point function using our recursion. While this integral is solvable by conventional methods, its simplicity allows us to illustrate all essential steps of our developed framework in a computationally tractable setting. The same procedure applies to higher \(N\)-point functions, although the complexity of the expressions increases significantly. Its low dimensionality makes it also possible to visualise the change in geometry during the limiting procedure. This is the contents of section~\ref{app:geometric_transition}.

\subsubsection{\(O(\eps)\) term}
According to our central recursion relation \eqref{eq:recursion}, the first-order \(\eps\) term \(\mathcal{I}_2^0\)
is determined by the finite part of the triangle's \(\mathcal{O}(1)\) term, \(\mathcal{I}_3^{-1}(\check{\cQ})\), in the limit where \(\cQ_{3,3} \to 0\). 
For brevity we use determinant-stripped, pure expressions \(\mathcal{J}_{N,d}^k(\cQ)\), defined by
\begin{equation}
    \mathcal{J}_{N,d}^{k} ({\cQ}) = \begin{cases}
     \sqrt{| \det {\cQ}_\textnormal{ext}|}  ~\mathcal{I}_{N,d}^{k} ({\cQ})\, , & N \textnormal{ odd}, \\[1ex]
      \sqrt{| \det {\cQ}|}  ~\mathcal{I}_{N,d}^{k} ({\cQ})\, , & N \textnormal{ even}.
    \end{cases}
\end{equation}
Then the triangle contribution is
given by (see ref.~\cite{Duhr:2025azh}):
\begin{equation}\label{eq:TriQQVol}
    \mathcal{J}_{3,4}^{-1} (\check{\cQ})
    = \frac{1}{8}\, A_{x,y} \left\{- G(1,y;x) + \left( G(1;x)-G(1;y) \right) G( y ; x) \right\}\, ,
\end{equation}
where we introduced the arguments
\begin{align}\label{eq:xy_Tri}
    x= \sqrt{\frac{\mathcal{Q}_{2,2}-\mathcal{Q}_{1,1}}{\mathcal{Q}_{2,2}}},\quad \textnormal{and} \quad y=\sqrt{\frac{\mathcal{Q}_{2,2}-\mathcal{Q}_{3,3}}{\mathcal{Q}_{2,2}}}\, ,
\end{align}
and the antisymmetrization operator \(A_{x,y}\) defined by
\beq
    A_{x} f(x) = f(x) - f(-x) \quad\textnormal{and}\quad A_{x,y} = A_x A_y = A_y A_x \, .
\eeq

To evaluate the limit \(\cQ_{3,3} \to 0\), we expand the argument \(y\) for small \(\cQ_{3,3}\). This corresponds to \(y \to 1\), which is a singular point. We isolate this singularity by expanding \(y\) around \(1\), i.e.,
\begin{equation}\label{eq:y=1-xi}
    y = \sqrt{\frac{\cQ_{2,2}-\cQ_{3,3}}{\cQ_{2,2}}} = 1 - \frac{\cQ_{3,3}}{2 \cQ_{2,2}} + \mathcal{O}\left( \left(\tfrac{\cQ_{3,3}}{\cQ_{2,2}}\right)^2  \right)\, .%
\end{equation}
The computational procedure involves two key steps: first bring the expression into a fibration basis with respect to 
\(\cQ_{3,3}\), effectively extracting the logarithmic singularities, and then apply the subtraction terms from eq.~\eqref{eq:divergencies} rendering the result finite. At \(\mathcal{O}(\eps)\) this is equivalent to simply setting the logarithmic singularities to zero, since finite term in eq.~\eqref{eq:divergencies} only appear starting \(\mathcal{O}(\eps^2)\) as we will see in section~\ref{sec:bub_eps2}.
The resulting finite expression for the \(\mathcal{O}(\eps)\) term is given by:
\begin{equation}
    \begin{split}
        \mathcal{J}^0_2 (\cQ ) &= \frac{1}{2}\Big[ G(1,1;x) + G(1,-1;x) - G(-1,1;x) - G(-1,-1;x) \\
        &\qquad+ G(0;\cQ_{2,2}) \left( G(1;x) - G(1;-x) \right)\Big]\, .
    \end{split}
\end{equation}
Beyond the computational role of the limit \(\cQ_{3,3}\to 0\) it also has a geometric interpretation: it realises a geometric transition from hyperbolic to Anti-de Sitter geometry at the level of the underlying orthoscheme.

\subsubsection{Geometric interpretation}\label{app:geometric_transition}
This interlude connects the computational framework developed above to a geometric picture. The \(\mathcal{O}(\eps)\) bubble result just obtained is the lowest-dimensional case in which the transition can be drawn explicitly, and we use it as our visual guide; the qualitative picture extends to all \(N\)-point orthoschemes.

\paragraph{Signature and geometric transition:} The key mathematical observation is that the signature of \(\check{\Q}_\textnormal{ext}\)
changes when crossing from \(\cQ_{N,N} < \cQ_{N+1,N+1} \) to \(\cQ_{N+1,N+1} < \cQ_{N,N} \). 
Then the signature of \(\check{\Q}_\textnormal{ext}\) changes from \((1,N+1)\) (hyperbolic) to \((2,N)\) (AdS) as \(\Q_{N+1,N+1}\) crosses \(\Q_{N,N}\) from above. The intermediate case \(\Q_{N+1,N+1} = \Q_{N,N}\) is degenerate geometry referred to as half-pipe geometry (see ref.~\cite{danciger_geometric_2013}). The signature stays \((2,N)\) as \(Q_{N,N}>\Q_{N+1,N+1} \to 0^+\). Geometric transitions like these have been studied in the mathematical literature, e.g., from hyperbolic to spherical geometry \cite{hodgson_degeneration_1986,porti_regenerating_1998} or more recently from hyperbolic to AdS geometry \cite{danciger_geometric_2011,danciger_geometric_2013,riolo_geometric_2022}. From the change in signature we already see that our geometric transition will be of the latter type.

\paragraph{Visualising the transition:}
We will now visualise this transition for the bubble Feynman integral. Since we take the \(\cQ_{3,3}\to 0\) limit of the triangle orthoscheme we start with a singly asymptotic three-dimensional orthoscheme in hyperbolic space. We need to understand the geometric transition this simplex undergoes as $\cQ_{3,3}$ varies. Setting $m_i^2 \coloneqq \cQ_{i,i}$, the hyperbolic regime corresponds to \(m_2<m_3\) while \(m_2 > m_3\) corresponds to AdS as we will see below: 

\paragraph{Hyperbolic regime \(m_2<m_3\):}
We use the Klein (projective) model of hyperbolic space, briefly introduced in section~\ref{sec:hyperbolic_volume}.
The hyperbolic tetrahedron can be described by four vertex vectors in the unit ball \(B_1(0)\subset\mathbb R^3\), equipped with the bilinear form \(\langle p,q\rangle=1-p\cdot q\), where \(p\cdot q\) is the standard scalar product on \(\mathbb R^3\) (cf.~eq.~\eqref{eq:Scalar_Product_Klein}).
The choice of vertices that reproduce the desired Gram matrix \(\check{\cQ}_\textnormal{ext}\) is not unique. One possible choice is
\begin{equation}\label{eq:klein_model_vectors}
    l_1 = \frac{1}{m_3}\begin{pmatrix} \sqrt{m_3^2 - m_2^2} \\ \sqrt{m_2^2 - m_1^2} \\ 0 \end{pmatrix}, \quad l_2= \frac{1}{m_3} \begin{pmatrix} \sqrt{m_3^2 - m_2^2} \\ 0 \\ 0 \end{pmatrix}, \quad l_3= \begin{pmatrix} 0 \\ 0 \\ 0 \end{pmatrix},\quad l_4= \begin{pmatrix} 0 \\ 0 \\ 1 \end{pmatrix}\, .
\end{equation}
The associated Gram matrix \(\bigl(\langle l_i,l_j\rangle\bigr)\) becomes
\begin{equation}\label{eq:gram_matrix_3dim}
    \check{\Q}_\textnormal{ext} = 
    \begin{pmatrix}
        \tfrac{m_1^2}{m_3^2} & \tfrac{m_2^2}{m_3^2} & 1 & 1 \\[1.3ex]
        \tfrac{m_2^2}{m_3^2} & \tfrac{m_2^2}{m_3^2} & 1 & 1 \\[1.3ex]
        1 & 1 & 1 & 1 \\
        1 & 1 & 1 & 0
    \end{pmatrix}\, .
\end{equation}
which is up to a volume-invariant rescaling the Gram matrix \(\check{\Q}_\textnormal{ext}\) as described in eq.~\eqref{eq:extended_Gram}.

\begin{figure}[ht!]
    \centering
    \begin{subfigure}{0.3\textwidth}
        \includegraphics[width=\linewidth]{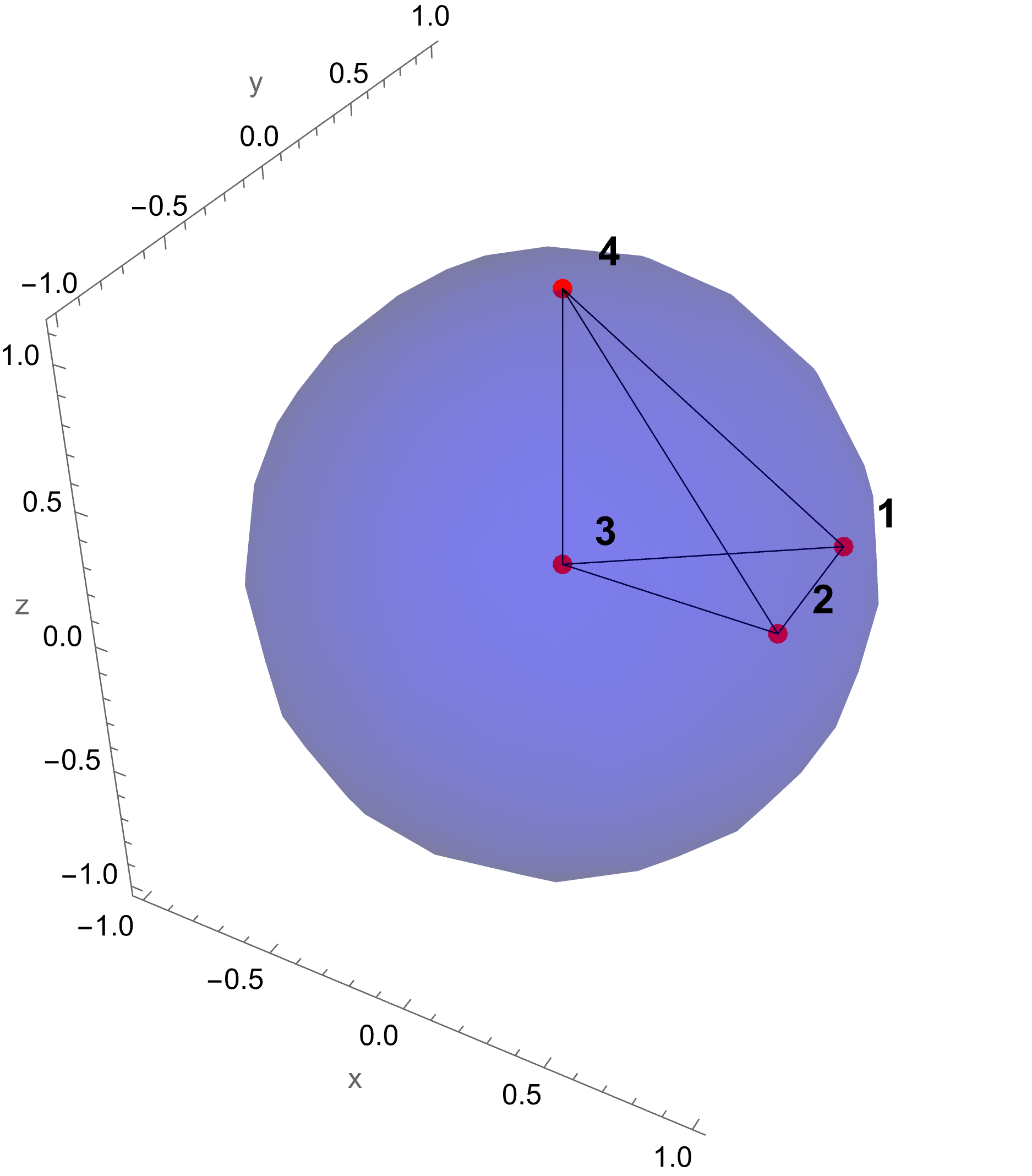}
        \caption{$m_3 = 1.5\,m_2$}
    \end{subfigure}
    \hfill
    \begin{subfigure}{0.3\textwidth}
        \includegraphics[width=\linewidth]{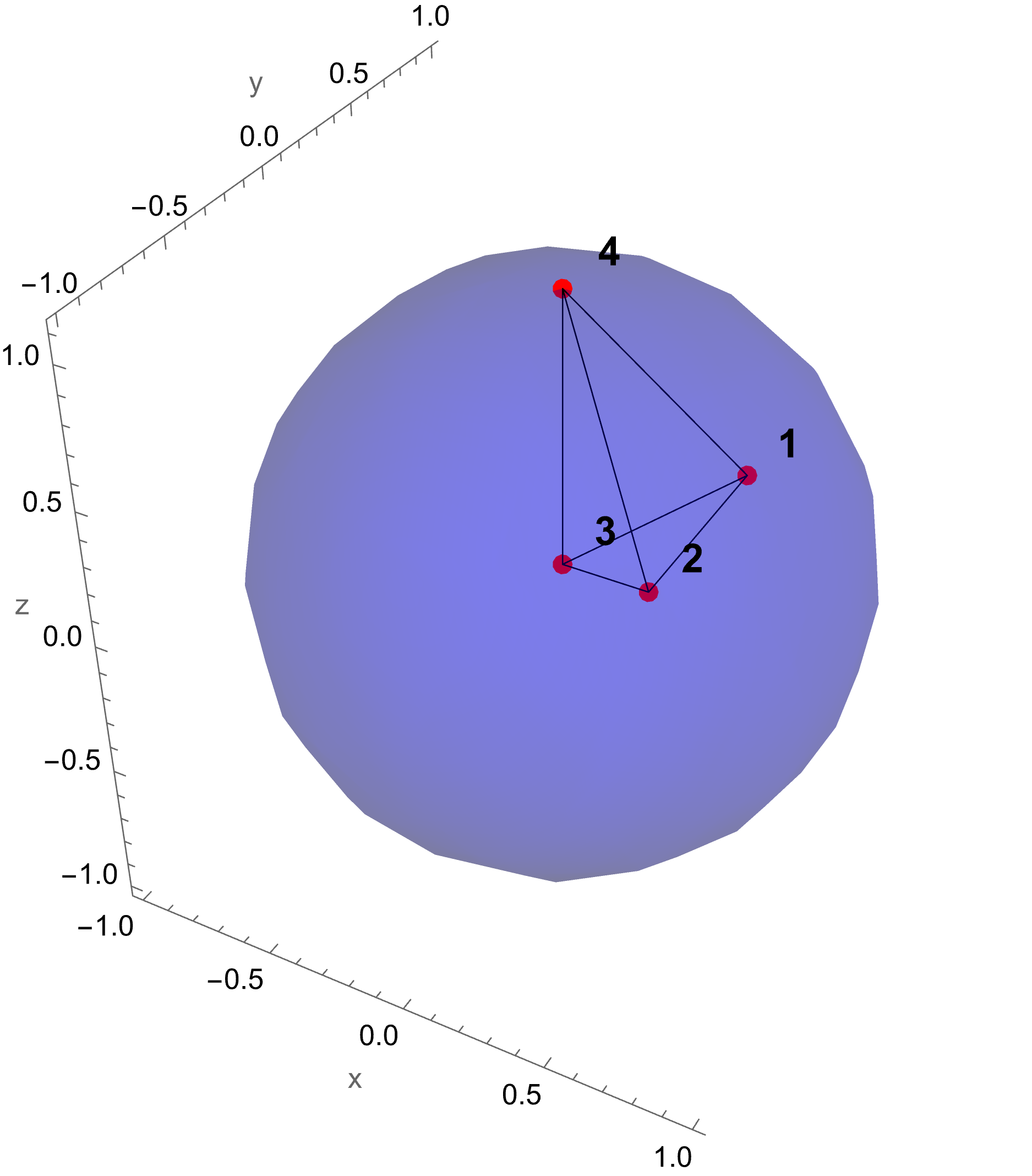}
        \caption{$m_3 = 1.05\,m_2$}
    \end{subfigure}
    \hfill
    \begin{subfigure}{0.3\textwidth}
        \includegraphics[width=\linewidth]{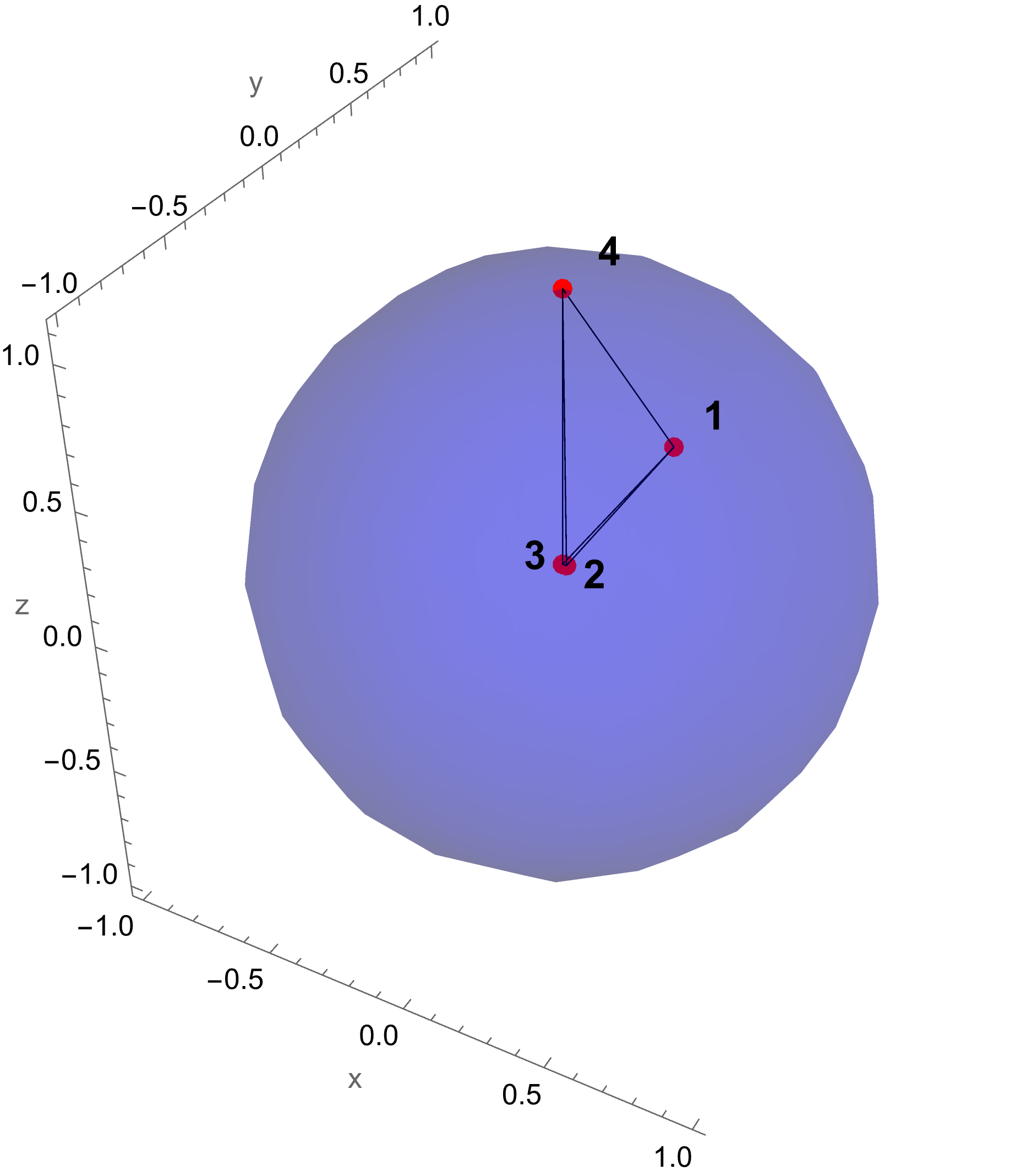}
        \caption{$m_3 = 1.0001\,m_2$}
        \label{fig:degenerate_tetrahedron}
    \end{subfigure}
    \caption{Comparison of configurations for different mass ratios.}
    \label{fig:mass-comparison}
\end{figure}

We have drawn the vectors for different mass configurations of \(m_3\) with \(m_2 < m_3\) in figure~\ref{fig:mass-comparison}. One can observe that the tetrahedron in the Klein model has vertices 2 and 3 approaching each other as \(m_3 \to m_2^+\). This leaves a degenerate tetrahedron (a triangle) for \(m_3 = m_2\) with hyperbolic volume zero (see figure~\ref{fig:degenerate_tetrahedron}).
Varying \(m_3\) below \(m_2\) introduces imaginary units in eq.~\eqref{eq:klein_model_vectors} through the square root \(\sqrt{m_3^2-m_2^2}\), necessitating a different approach.

\paragraph{Crossing the threshold into AdS regime \(m_3 < m_2\):} We can make the transition visible by letting the ambient bilinear form vary with respect to the mass parameters and especially with respect to \(m_3\) instead of only plugging values into the vectors. For this we define the generalised bilinear form
\begin{equation}
    d_\Lambda (p,q) = 1 - ( \Lambda p_1 q_1 + p_2 q_2 + p_3 q_3 + \ldots).
\end{equation}
One possibility is to pick \(\Lambda = 1- \tfrac{m_2^2}{m_3^2}\) then our Gram matrix \eqref{eq:gram_matrix_3dim} stays the same when we change our vectors \eqref{eq:klein_model_vectors} accordingly, i.e.,
\begin{equation}\label{eq:klein_model_vectors_prime}
    l^\prime_1 = \begin{pmatrix} 1 \\ \tfrac{\sqrt{m_2^2 - m_1^2}}{m_3}  \\ 0 \end{pmatrix}, \quad l^\prime_2= \begin{pmatrix} 1 \\ 0 \\ 0 \end{pmatrix}, \quad l^\prime_3= \begin{pmatrix} 0 \\ 0 \\ 0 \end{pmatrix},\quad l^\prime_4= \begin{pmatrix} 0 \\ 0 \\ 1 \end{pmatrix}.
\end{equation}
Changing the bilinear form correspondingly modifies the defining quadric of the model.

\begin{figure}[t]
    \centering
    \begin{subfigure}{0.3\textwidth}
        \includegraphics[width=\linewidth]{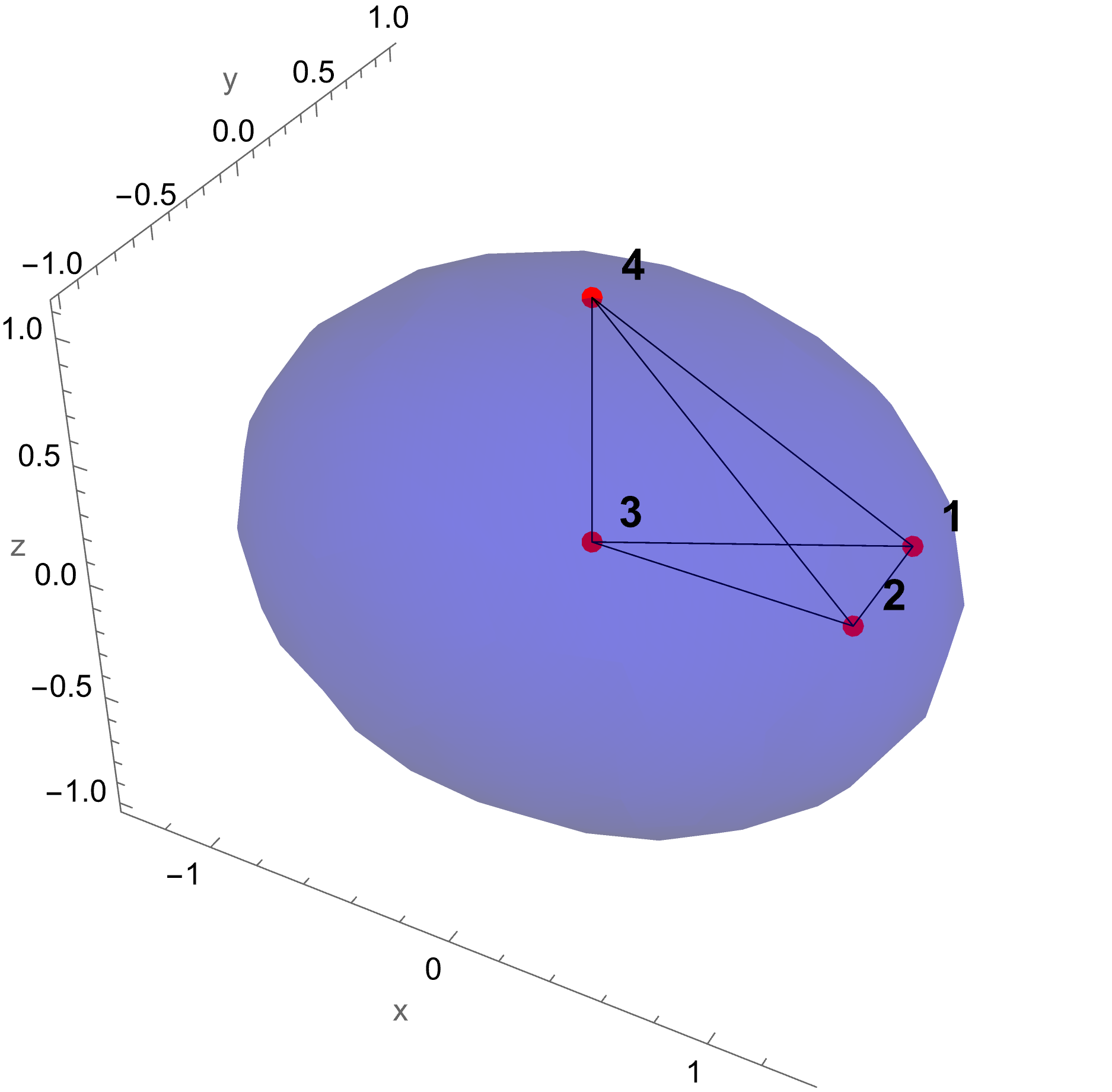}
        \caption{$m_3 = 1.5\,m_2$}
        \label{fig:hyperbolic1dot5Lambda}
    \end{subfigure}
    \hfill
    \begin{subfigure}{0.3\textwidth}
        \includegraphics[width=\linewidth]{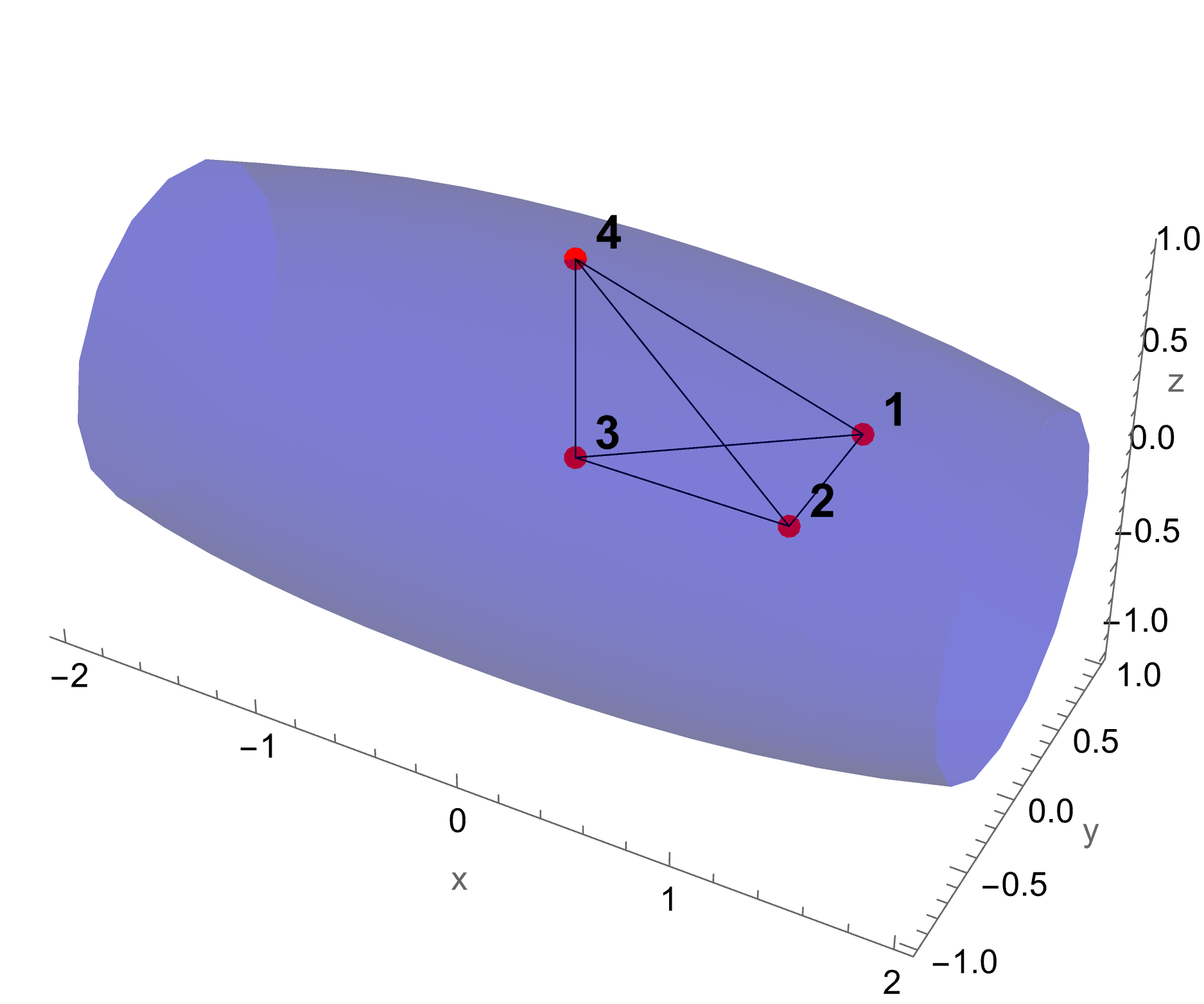}
        \caption{$m_3 = 1.05\,m_2$}
    \end{subfigure}
    \hfill
    \begin{subfigure}{0.3\textwidth}
        \includegraphics[width=\linewidth]{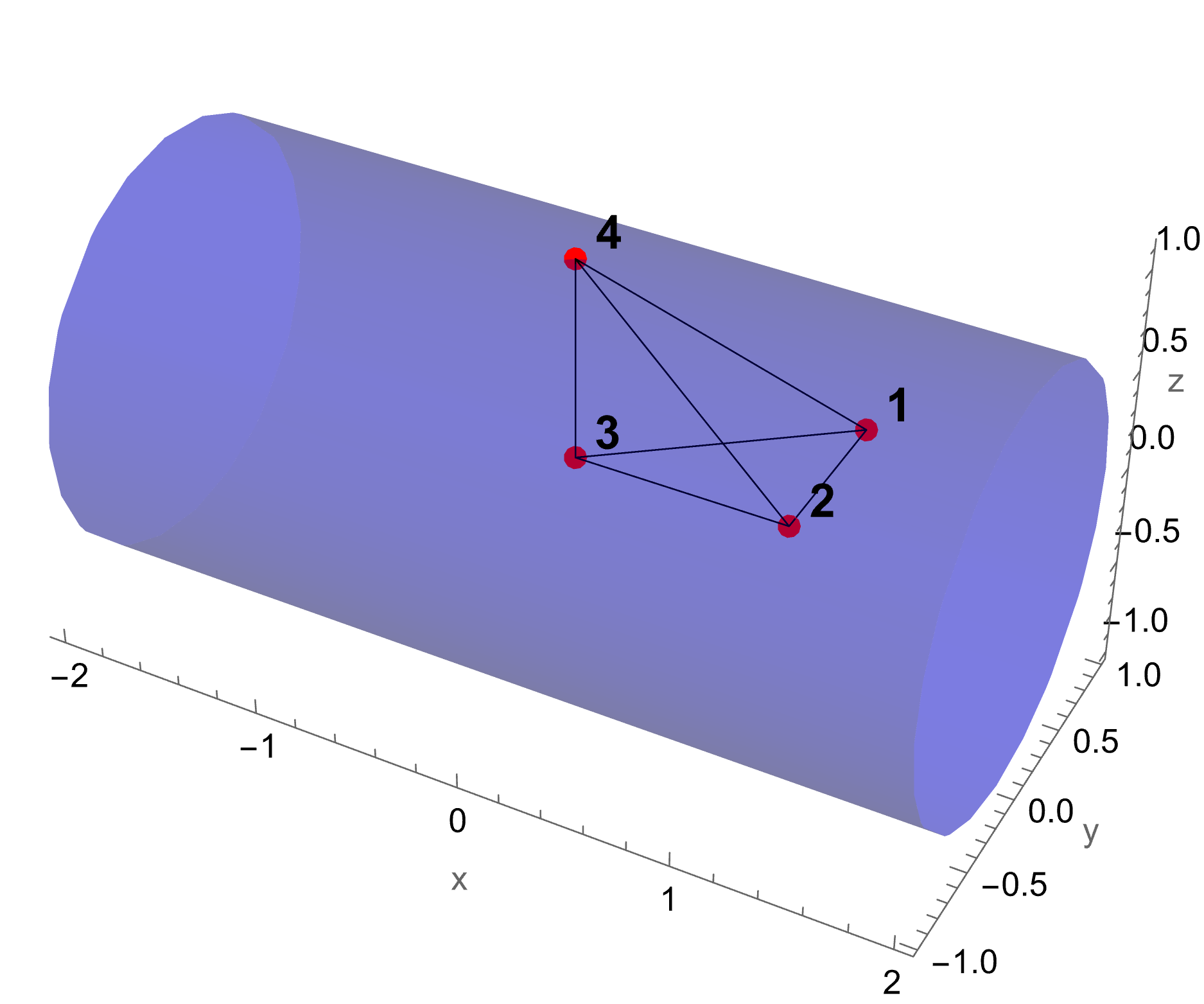}
        \caption{$m_3 = m_2$}
        \label{fig:hyperbolic1dot0001Lambda}
    \end{subfigure}
    \begin{subfigure}{0.3\textwidth}
        \includegraphics[width=\linewidth]{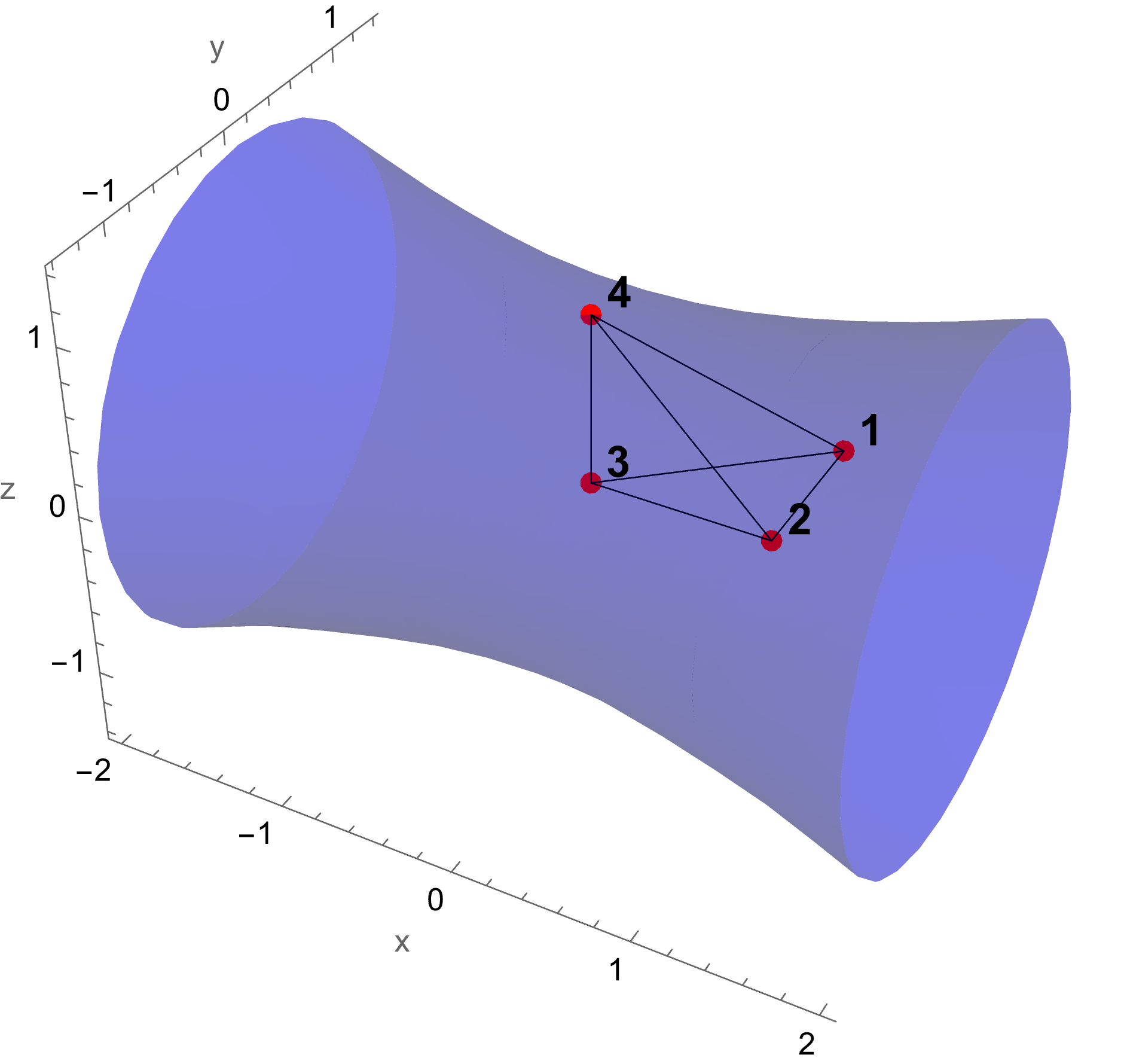}
        \caption{$m_3 = 0.9\,m_2$}
        \label{fig:AdS0dot9Lambda}
    \end{subfigure}
    \hspace*{5mm}
    \begin{subfigure}{0.3\textwidth}
        \includegraphics[width=\linewidth]{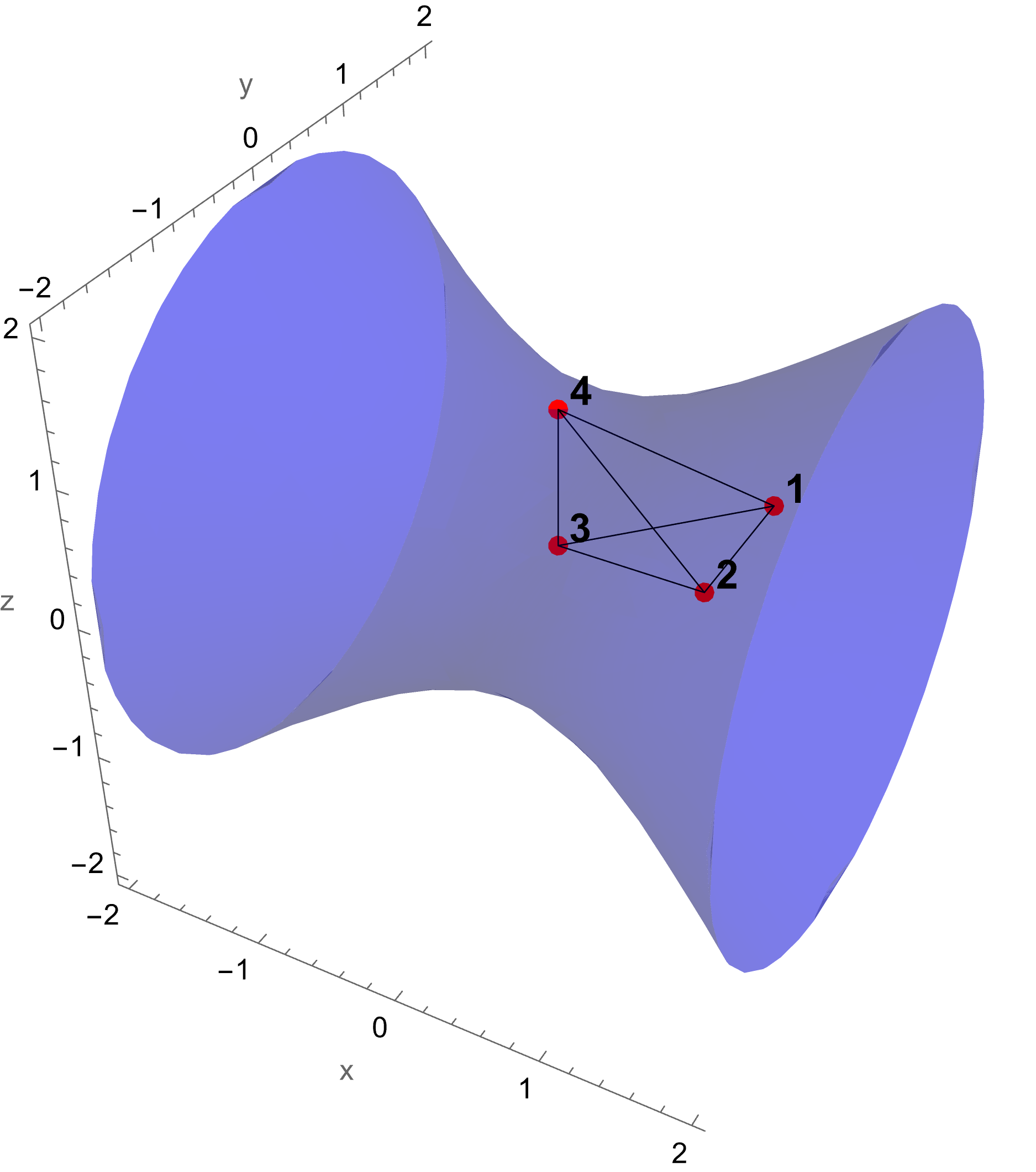}
        \caption{$m_3 = 0.75\,m_2$}
        \label{fig:AdS0dot75Lambda}
    \end{subfigure}
    \caption{Geometric transition from hyperbolic to AdS geometry for different mass ratios $m_3/m_2$. The quadric changes from a sphere ($m_3 > m_2$, hyperbolic regime) through a cylinder ($m_3 = m_2$, half-pipe geometry) to a one-sheeted hyperboloid ($m_3 < m_2$, AdS regime).}
    \label{fig:mass-comparison_beyond}
\end{figure}

With this new approach we can draw images also for \(m_3 < m_2\) within the Klein model of \(\textnormal{AdS}_3\) (see figs.~\ref{fig:AdS0dot9Lambda} and~\ref{fig:AdS0dot75Lambda}); note also the elongation of the Klein sphere in progress from figs.~\ref{fig:hyperbolic1dot5Lambda} to~\ref{fig:hyperbolic1dot0001Lambda}. The main takeaway here is that the quadric changes from a sphere to a cylinder (degenerate case with one direction having eigenvalue zero) to the one-sheeted hyperboloid as known from AdS geometry.
\begin{figure}[htb!]
    \centering
    \begin{subfigure}{0.3\textwidth}
        \includegraphics[width=\linewidth]{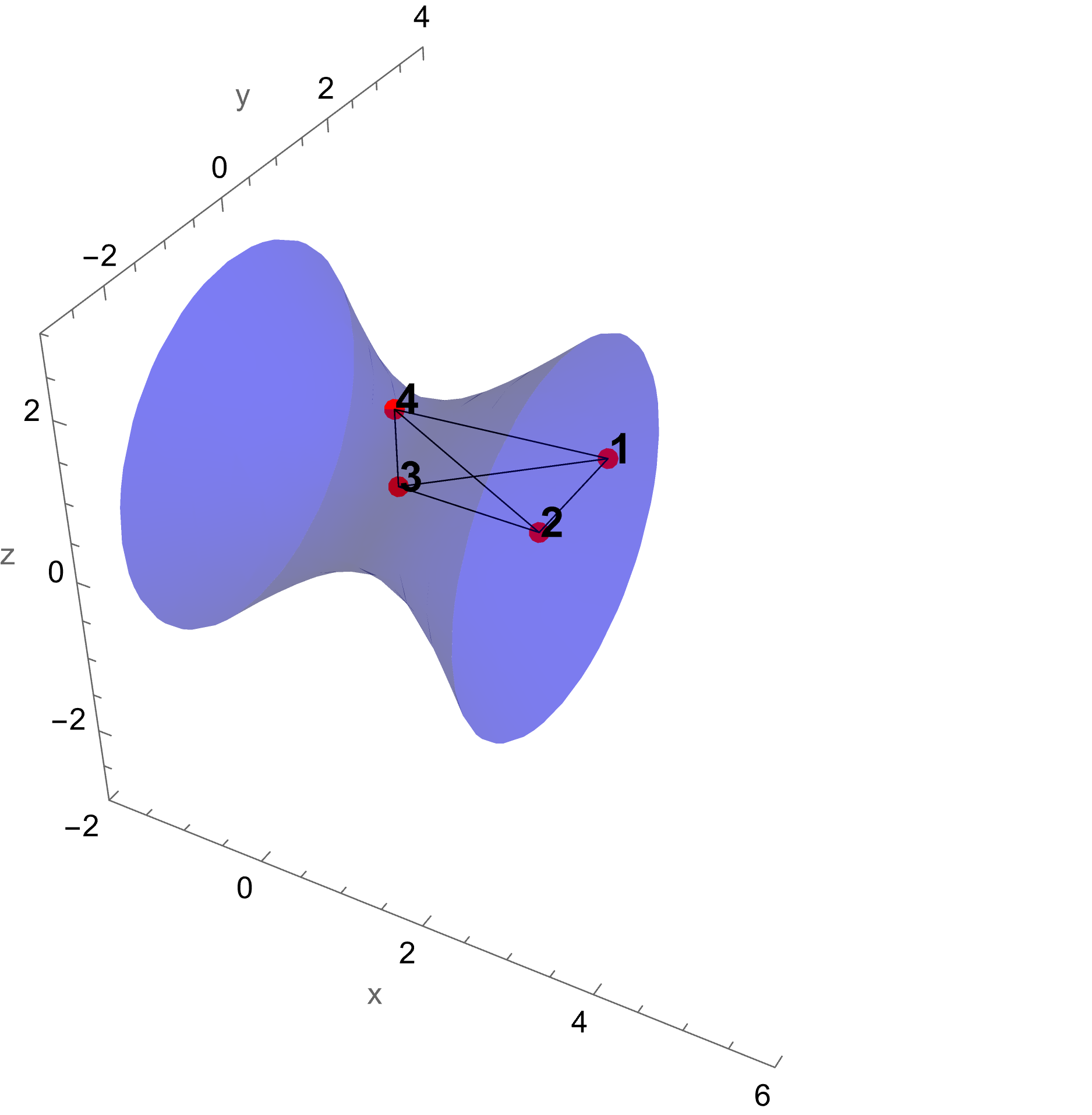}
        \caption{$m_3 = 0.5\,m_2$}
    \end{subfigure}
    \hfill
    \begin{subfigure}{0.3\textwidth}
        \includegraphics[width=\linewidth]{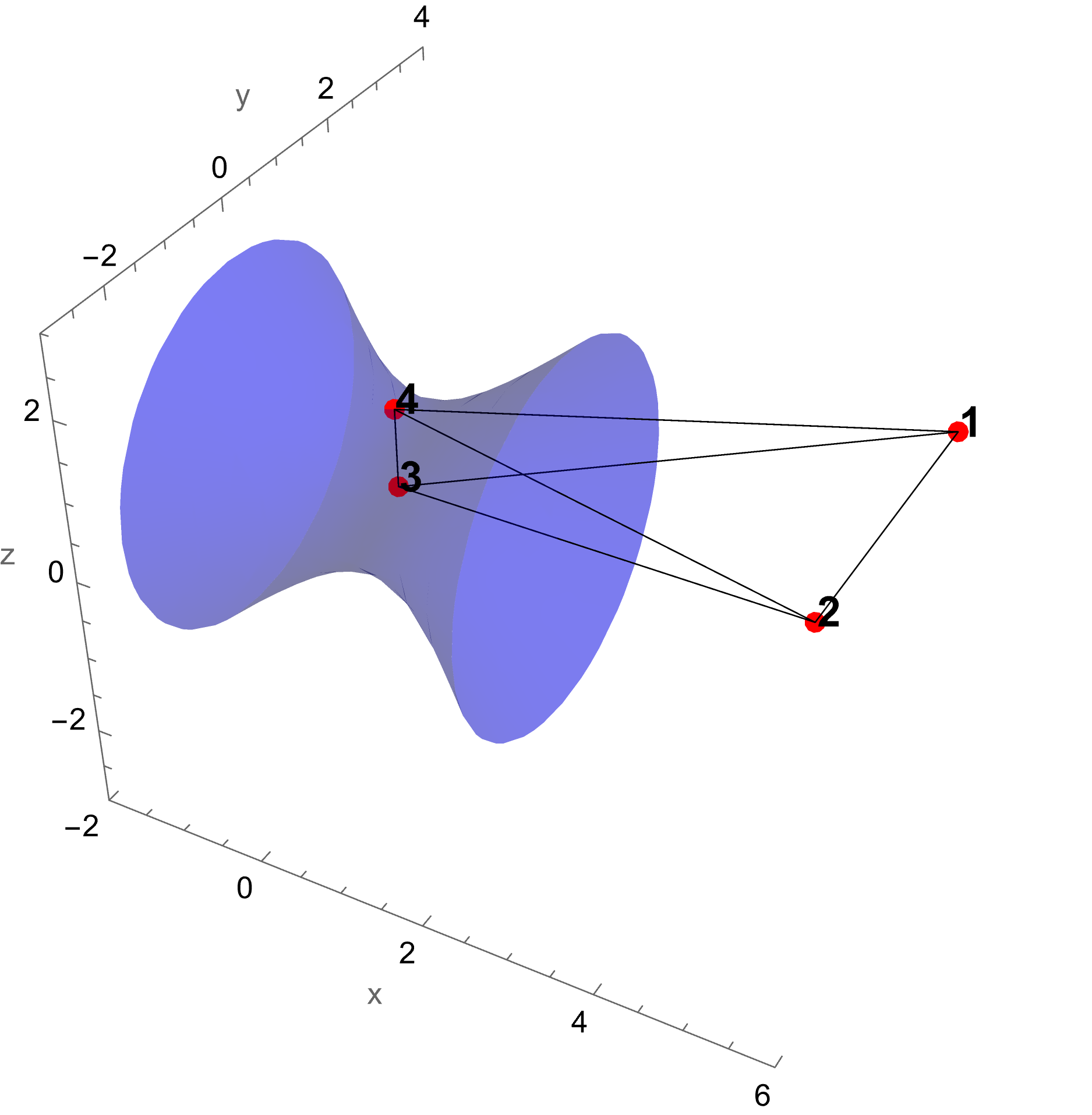}
        \caption{$m_3 = 0.2\,m_2$}
    \end{subfigure}
    \hfill
    \begin{subfigure}{0.3\textwidth}
        \includegraphics[width=\linewidth]{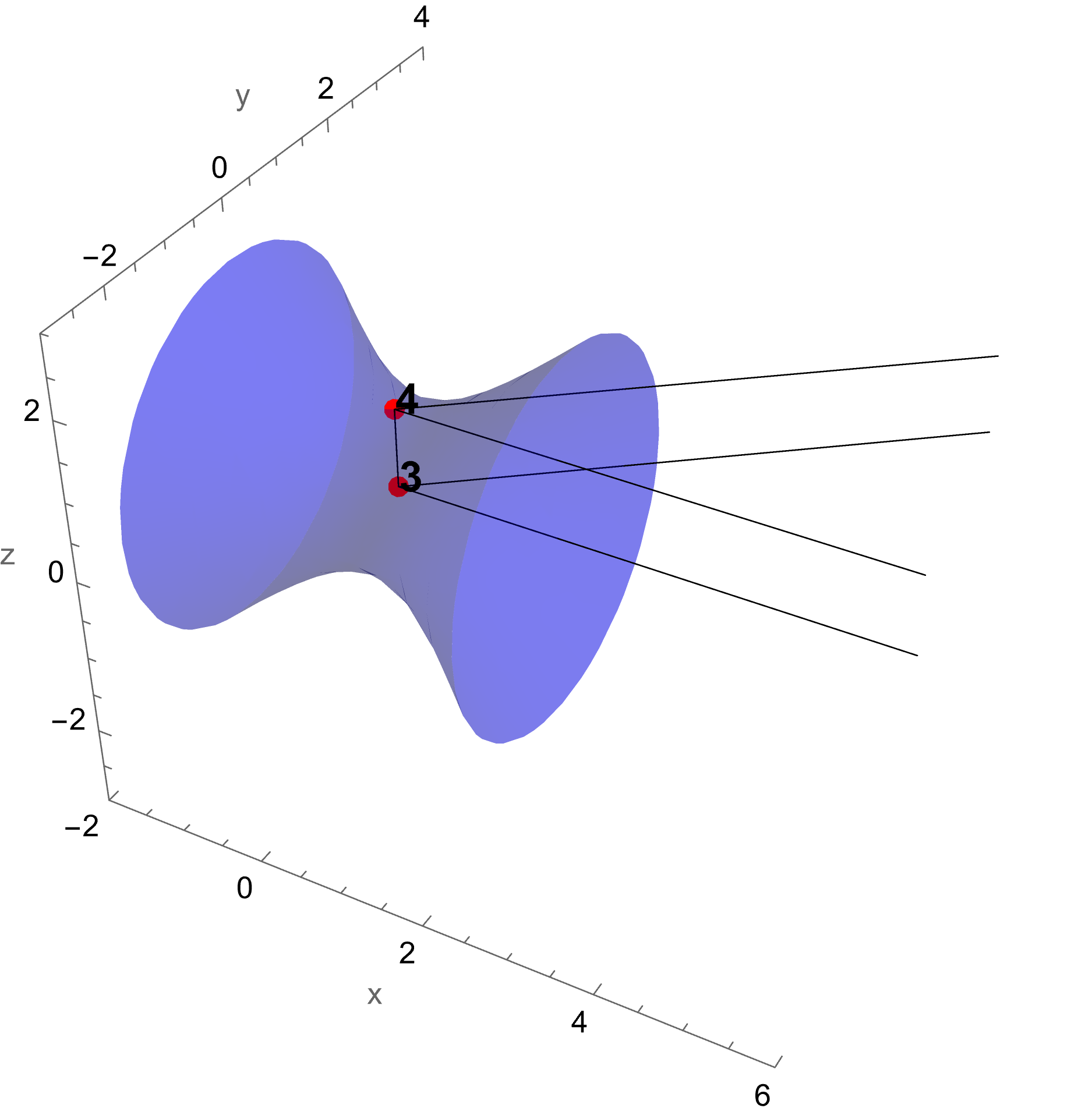}
        \caption{$m_3 = 0.0001\,m_2$}
    \end{subfigure}
    \caption{Tetrahedron in the Klein model of AdS space as $m_3 \to 0^+$. Vertices 1 and 2 escape to infinity, and the divergent geodesic separation between faces $(12)$ and $(34)$ corresponds to the logarithmic term subtracted in the recursion.}
    \label{fig:mass-comparison_m_3->0}
\end{figure}
\paragraph{Limit \(m_3 \to 0^+\):} Since the shape of the tetrahedron in figure~\ref{fig:mass-comparison_beyond} only changes marginally with the varying bilinear form \(d_\Lambda\), it is more instructive to fix the bilinear form at \(d_{-1}\) and  again let the vectors evolve, i.e.,
\begin{equation}\label{eq:klein_model_vectors_AdS}
    l_1 = \frac{1}{m_3}\begin{pmatrix} \sqrt{m_2^2 - m_3^2} \\ \sqrt{m_2^2 - m_1^2} \\ 0 \end{pmatrix}, \quad l_2= \frac{1}{m_3} \begin{pmatrix} \sqrt{m_2^2 - m_3^2} \\ 0 \\ 0 \end{pmatrix}, \quad l_3= \begin{pmatrix} 0 \\ 0 \\ 0 \end{pmatrix},\quad l_4= \begin{pmatrix} 0 \\ 0 \\ 1 \end{pmatrix} \, .
\end{equation}
In figure~\ref{fig:mass-comparison_m_3->0} we can see that as \(m_3 \to 0^+ \) vertices one and two are escaping to infinity. 
The geodesic separation between the faces $(12)$ and $(34)$ diverges logarithmically in this limit. This divergent geodesic length is precisely the term that is subtracted in the recursive construction to isolate the finite order $\varepsilon$ terms of the bubble.
Recall that the hyperbolic length of the edge \((12)\) is the one-dimensional volume associated to the bubble at \(\varepsilon = 0\).
This completes the geometric picture for the \(\mathcal{O}(\eps)\) bubble term.

More generally, the recursive structure suggests a sequence of geometric transitions at each order in $\varepsilon$. Descending the ladder $\cQ_{N+1,N+1} \to 0$, $\cQ_{N,N} \to \infty$, $\cQ_{N-1,N-1} \to 0$, $\ldots$, the signature appears to change as $(1,N+1) \to (2,N) \to (2,N) \to (3,N-1) \to \cdots$, where the odd steps $\cQ_{2k+1,2k+1} \to 0$ increment the number of timelike directions while the even steps $\cQ_{2k,2k} \to \infty$ leave the signature unchanged. We leave a detailed geometric analysis of these higher transitions to future work. 
\subsubsection{\(O(\eps^2)\) term}\label{sec:bub_eps2}
The \(O(\eps^2)\) term of the bubble diagram, \(\mathcal{I}_2^1(\cQ)\), is determined by the finite part of the triangle's \(\mathcal{O} (\eps) \) term, \(\mathcal{I}_{3,4}^0(\check{\cQ})\), in the same \( {\cQ}_{33} \to 0\) limit. 
Although one could derive $\mathcal{I}_{3,4}^0(\check{\Q})$ directly from $\mathcal{I}_{5,6}^{-1}(\check{\Q})$ using both eqs.~\eqref{eq:eventoodd} and \eqref{eq:recursion}, the pentagon expression is already lengthy and would unnecessarily obscure the present illustration.
Instead we simply use the available result for \(\mathcal{I}_{3,4}^0(\check{\cQ})\) given in ref.~\cite{Duhr:2025azh}:
\begin{equation}\bsp
        \mathcal{J}_{3,4}^{0} (\check{\cQ})= \frac{1}{8} A_{x,y} \bigg\{& G(y,1,1;x) + G(y,1,-1;x) - \left( G(1,1;y)-G(1,-1;y) \right)G(y;x)  \\
        & +\log (\cQ_{2,2}) \left( -G(1,y;x) + (G(1;x) -G(1;y) ) G(y;x) \right) \bigg\}\, .
\esp\end{equation}

Applying the same procedure as for the \(\mathcal{O}(\eps)\) term we obtain the result:
\begin{equation}
    \begin{split}
        \mathcal{J}_2^1 (\cQ) &= \frac{1}{2} \Bigg[ \sum_{s_1,s_2,s_3 =\pm1} s_1 \, G(s_1,s_2,s_3;x) + \log \cQ_{2,2} \sum_{s_1,s_2 =\pm1} s_1\, G(s_1,s_2;x) \\
        &\qquad + \left( \frac{(\log Q_{2,2})^2}{2} - \zeta_2 \right) \sum_{s_1 =\pm1}\, s_1\, G(s_1;x) \Bigg] \, .
    \end{split}
\end{equation}
This example, while simple, demonstrates the power and systematic nature of the geometric recursion introduced in this paper. The higher order terms in \(\eps\) for any \(N\)-point function can be obtained by applying this limiting procedure recursively as depicted in figure~\ref{fig:tikz_flowchart}. 

\subsection{The box at \(\mathcal{O}(\eps)\)}\label{app:boxeps}
The $\mathcal{O}(\eps)$ coefficient of the box integral was already computed in ref.~\cite{Duhr:2025azh}. 
However, the representation obtained in ref.~\cite{Duhr:2025azh} relied on rationalising two independent square roots, leading to polylogarithmic expressions with nested square-root arguments.
In contrast, our geometric recursion avoids any rationalisation procedure. All polylogarithms appear with single square root arguments, and no nested square roots occur. 
Since our recursion in eq.~\eqref{eq:recursion} writes the \(\mathcal{O}(\eps)\) contribution as a limit of the integer-dimensional pentagon integral, some of its 
structure carries over. In particular, applying the recursion and taking the limit \(\cQ_{5,5} \to 0\), we obtain:
\beq\bsp
\mathcal{J}_4^0(\cQ)
=
\frac{1}{8}&\Big(
A  _{x_1,y_1}(F_1)
+
A  _{x_2,y_2}(F_2)
+
A  _{x_3,y_3}(F_3)
+
A  _{x_1,y_1,s}(F_4)
+
A  _{x_2,y_2,s}(F_5)\\
&\quad
+
A  _{x_3,y_3,s}(F_6)
\Big) \Big|_{s=1}  - \log (\cQ_{4,4})\, \mathcal{J}^{-1}_4 (\cQ) \, .
\esp\eeq
The contributions \(F_2\) and \(F_3\) in here come from \(\Li_{1,2}\) terms in Rudenko's formula (see eq.~4.34 in ref.~\cite{Duhr:2025azh} for details), while \(F_1\) comes from combining logarithms in \(\cQ_{2,2}\) and \(\cQ_{4,4}\) together. They are explicitly given by:
\begin{align}
F_1 &= \left(G(1; x_2) + G(-1; x_2) - G(1; x_1) - G(-1;x_1)\right)
\, G(1; x_1)\,G(1; y_1) \, ,
\\[0.5em]
F_2 &= G(0; y_2)\,G(1; x_2)\,G(1; y_2)
+2\,G(0; |x_2|)\,G(1; x_2)\,G(y_2; x_2)
\nonumber\\
&\quad -2\,G(0; y_2)\,G(1; x_2)\,G(y_2; x_2)
+G(0; y_2)\,G(1; y_2)\,G(y_2; x_2)
\nonumber\\
&\quad -2\,G(1; x_2)\,G(0,y_2; x_2)
+G(0; y_2)\,G(1,y_2; x_2)
\nonumber\\
&\quad +G(0; -y_2)\Big(
-2\,G(1; x_2)\,G(y_2; x_2)
+G(1; y_2)\big(G(1; x_2)+G(y_2; x_2)\big)
+G(1,y_2; x_2)
\Big)
\nonumber\\
&\quad +2\,G(1,0,y_2; x_2)\, ,
\\[0.5em]
F_3 &= -\,G(0; y_3)\,G(1; x_3)\,G(1; y_3)
-2\,G(0; |x_3|)\,G(1; x_3)\,G(y_3; x_3)
\nonumber\\
&\quad +2\,G(0; y_3)\,G(1; x_3)\,G(y_3; x_3)
-G(0; y_3)\,G(1; y_3)\,G(y_3; x_3)
+2\,G(1; x_3)\,G(0,y_3; x_3)
\nonumber\\
&\quad +G(0; -y_3)\Big(
2\,G(1; x_3)\,G(y_3; x_3)
- G(1; y_3)\big(G(1; x_3)+G(y_3; x_3)\big)
- G(1,y_3; x_3)
\Big)
\nonumber\\
&\quad -G(0; y_3)\,G(1,y_3; x_3)
-2\,G(1,0,y_3; x_3)\, .
\end{align}
The functions \(F_4,F_5,F_6\) on the other hand depend on \(s\) which comes, similarly as for the bubble, from arguments \(\sqrt{1-\tfrac{\cQ_{5,5}}{\cQ_{i,i}}}\) of the pentagon tending to \(1\) in the \(\cQ_{5,5}\) limit. These functions are given by:
\begin{align}
F_4 =\;&
- G(0; x_3 y_1)\,G(1; x_1)\,G(1; s y_1)
+ G(1; x_1)\,G(1; s y_1)\,G(1; x_3 y_1)
\notag\\
&- G(1; x_1)\,G(1; s y_1)\,G(s; x_3)
+ G(0; x_3 y_1)\,G(1; x_1)\,G(s y_1; x_1)
\notag\\
&- G(1; x_1)\,G(1; y_1)\,G(s y_1; x_1)
- G(0; x_3 y_1)\,G(1; s y_1)\,G(s y_1; x_1)
\notag\\
&- G(1; x_1)\,G(1; x_3 y_1)\,G(s y_1; x_1)
+ G(1; s y_1)\,G(1; x_3 y_1)\,G(s y_1; x_1)
\notag\\
&+ G(1; x_1)\,G(s; x_1)\,G(s y_1; x_1)
+ 2\,G(1; x_1)\,G(s; x_3)\,G(s y_1; x_1)
\notag\\
&- G(1; s y_1)\,G(s; x_3)\,G(s y_1; x_1)
- G(1; x_3 y_1)\,G(s; x_3)\,G(s y_1; x_1)
\notag\\
&+ G(1; x_1)\,G(y_1; x_1)\,G(s y_1; x_1)
- G(1; y_1)\,G(y_1; x_1)\,G(s y_1; x_1)
\notag\\
&- G(1; x_1)\,G(s y_1; x_1)\,G(x_3 y_1; x_1)
+ G(1; x_3 y_1)\,G(s y_1; x_1)\,G(x_3 y_1; x_1)
\notag\\
&-\big(G(0; x_3 y_1)-G(1; x_3 y_1)+G(s; x_3)\big)\,
G(1,s y_1; x_1)
\notag\\
&+ G(0; x_3) G(1; x_1)\big(G(1; s y_1)-2\,G(s y_1; x_1)\big)
+G(0; x_3) G(1,s y_1; x_1)
\notag\\
&+G(0; x_3) \big(G(1; s y_1)+G(1; x_3 y_1)\big)\,G(s y_1; x_1) 
- G(0; y_1)G(1; x_1)\,G(s y_1; x_1)
\notag\\
&+ G(0; y_1) G(1; s y_1)\big(G(1; x_1)+G(s y_1; x_1)\big)
+ G(0; y_1) G(1,s y_1; x_1)
\notag\\
&- G(1; x_1)\,G(s,s y_1; x_1)
- G(1; x_1)\,G(y_1,s y_1; x_1)
+ G(1; y_1)\,G(y_1,s y_1; x_1)
\notag\\
&+ G(1; x_1)\,G(x_3 y_1,s y_1; x_1)
- G(1; x_3 y_1)\,G(x_3 y_1,s y_1; x_1)
\notag\\
&+ G(1,s,s y_1; x_1)
+ G(1,y_1,s y_1; x_1)
- G(1,x_3 y_1,s y_1; x_1)\, ,
\displaybreak[1]\\[6pt]
F_5 =\;&
G(0; y_2)\Big(
G(1; x_2)\,G(s y_2; x_2)
- G(1; s y_2)\big(G(1; x_2)+G(s y_2; x_2)\big)
- G(1,s y_2; x_2)
\Big)
\notag\\
&+ G(1; y_2)\Big(
G(y_2; x_2)\,G(s y_2; x_2)
- G(y_2,s y_2; x_2)
\Big)
\notag\\
&+ G(1; x_2)\big(G(1; y_2)-G(s; x_2)-G(y_2; x_2)\big)\,G(s y_2; x_2)
+ G(1; x_2) G(s,s y_2; x_2)
\notag\\
&+ G(1; x_2) G(y_2,s y_2; x_2)
- G(1,s,s y_2; x_2)
- G(1,y_2,s y_2; x_2)\, ,
\displaybreak[0]\\[6pt]
F_6 =\;&
G(0; y_3)\Big(
- G(1; x_3)\,G(s y_3; x_3)
+ G(1; s y_3)\big(G(1; x_3)+G(s y_3; x_3)\big)
+ G(1,s y_3; x_3)
\Big)
\notag\\
&+ G(1; x_3) \big(-G(1; y_3)+G(s; x_3)+G(y_3; x_3)\big)\,G(s y_3; x_3)
- G(1; x_3) G(s,s y_3; x_3)
\notag\\
&- G(1; x_3) G(y_3,s y_3; x_3)
+ G(1; y_3)\Big(
- G(y_3; x_3)\,G(s y_3; x_3)
+ G(y_3,s y_3; x_3)
\Big)
\notag\\
&+ G(1,s,s y_3; x_3)
+ G(1,y_3,s y_3; x_3)\, .
\end{align}
While the arguments are given through:
\begin{equation}
    x_1 = \sqrt{\frac{\cQ_{2,2}-\cQ_{1,1}}{\cQ_{2,2}}}\,, \qquad
    x_2 = \sqrt{\frac{\cQ_{4,4}-\cQ_{1,1}}{\cQ_{4,4}}}\,, \qquad
    x_3 = \sqrt{\frac{\cQ_{4,4}-\cQ_{3,3}}{\cQ_{4,4}}}\,,
\end{equation}
and
\begin{equation}
    y_1 = \iu\,\sqrt{\frac{(\cQ_{3,3}-\cQ_{2,2})\,\cQ_{4,4}}{\cQ_{2,2}(\cQ_{4,4}-\cQ_{3,3})}}\,, \ 
    y_2 = \iu\,\sqrt{\frac{(\cQ_{3,3}-\cQ_{2,2})(\cQ_{4,4}-\cQ_{1,1})}{(\cQ_{2,2}-\cQ_{1,1})(\cQ_{4,4}-\cQ_{3,3})}}\,, \ 
    y_3 = \iu\,\sqrt{\frac{\cQ_{3,3}-\cQ_{2,2}}{\cQ_{2,2}-\cQ_{1,1}}}\,.
\end{equation}
In particular, all $x_i > 0$ and all $y_i$ are purely imaginary for generic 
Euclidean kinematics, i.e.\ $0 < \cQ_{1,1} < \cQ_{2,2} < 
\cQ_{3,3} < \cQ_{4,4}$.
\subsubsection{The triangle at \(\mathcal{O}(\eps)\)}
The triangle is in fact more involved than the box within our recursive framework, as it appears as a limiting case of the box integral.
We therefore expand the pure part of the box at $\mathcal{O}(\eps)$ around $\tfrac{1}{\cQ_{4,4}} = 0$.
This can similarly be done to the other limits by defining \(\xi \coloneqq \tfrac{1}{\cQ_{4,4}}\) and move to a fibration basis of \(\xi\). 
A useful consistency check is the vanishing of the coefficient of $G(0;\xi)$ in the limit $\xi \to 0$.
We obtain an expression slightly more complicated than the expression found in ref.~\cite{Duhr:2025azh}; however, we checked analytically that these expression coincide, validating our procedure.

Higher orders in \(\eps\) can be obtained analogously by considering higher-point integrals. In practice, however, the complexity grows rapidly. For example, the computation of the hexagon at $\mathcal{O}(\eps)$ in section~\ref{sec:hexagon} already constitutes a necessary intermediate step for determining the box and triangle at $\mathcal{O}(\eps^2)$. 
\section{Hexagon at \(\mathcal{O}(\varepsilon)\)}\label{sec:hexagon}
In ref.~\cite{Duhr:2025azh}, analytic results for dimensionally-regularised one-loop triangle, box, and pentagon integrals were obtained to higher orders in \(\eps\). Extending these results to higher-point integrals, such as the hexagon, can provide further insight into the general structure of one-loop functions and the function spaces appearing in integration-by-parts reductions.
In particular, the hexagon in $D=6$ dimensions appears in $\mathcal{N}=4$ super Yang--Mills theory, where it is related via differential equations to the two-loop remainder function~\cite{Dixon:2011ng}. 
Since higher orders in $\eps$ correspond to higher transcendental weight, similarly to higher-loop contributions, this may indicate a structural connection between the $\eps$-expansion of one-loop integrals and higher-loop amplitudes.

As a first non-trivial application of the recursion, we compute the coefficient \(\mathcal{I}_6^0(Q)\), corresponding to the \(\mathcal{O}(\eps)\) contribution of the one-loop six-point integral in \(D=6-2\eps\) dimensions. To our knowledge, this is the first explicit closed MPL expression for this coefficient with arbitrary masses and off-shell Euclidean kinematics. The massless on-shell hexagon at the same order was studied in ref.~\cite{Henn:2022ydo}.
We assume generic Euclidean kinematics in the following, i.e., we need all orthoschemes of the dissection to be well-defined. Results for other kinematic regimes can be obtained via systematic analytic continuation (see section~5 in ref.~\cite{Duhr:2025azh} for the analytic continuation in case of the one-loop triangle diagram). Likewise, non-generic kinematics can be reached via careful limiting procedures.

It suffices to calculate \(\mathcal{I}^0_{6} (\cQ)\) for \(\cQ \in \mathcal{G}_6^{\mathrm{nest}}\), since \(\mathcal{I}^0_{6} (Q)\) for general kinematics is obtained by orthoscheme dissection via eq.~\eqref{eq:Integral_Dissection}. According to the recursion relation~\eqref{eq:recursion}, \(\mathcal{I}^0_{6} (\cQ)\) is obtained from the \(\varepsilon = 0\) contribution of the heptagon in eight dimensions, \(\mathcal{I}^{-1}_{7,8} (\check{\cQ})\). The heptagon is a weight-four function~\cite{Rudenko2020Orthoschemes,Ren:2023tuj}; hence \(\mathcal{I}^0_{6} (\cQ)\) is a weight-four function as well, in accordance with eq.~\eqref{eq:weight_formula}.
Since the resulting expressions are lengthy, we collect them in the ancillary file \texttt{hexagon.wl}, which contains the analytic expression for \(\mathcal{I}_6^0(\cQ)\) (global determinant factor omitted) for \(\mathcal{Q} \in \mathcal{G}_6^{\mathrm{nest}}\) in terms of multiple polylogarithms in \texttt{PolyLogTools} notation~\cite{Duhr:2019tlz}. The kinematic variables are the diagonal entries \(\cQ_{i,i}\), \(i=1,\ldots,6\), of the nested Gram matrix. 
The full hexagon \(\mathcal{I}_6^0(Q)\) for general kinematics is recovered by summing the included expression over \(\mathcal{Q}(\mathbf{j}) \in \mathrm{BS}(Q)\) as in eq.~\eqref{eq:Integral_Dissection}, weighted by the sign factors of eq.~\eqref{eq:sgn_Factor}. We outline the main steps of the calculation below.

The MPL arguments appearing in $\cI_6^0(\cQ)$ are limits, in the $\cQ_{7,7} \to 0$ regime, of the arguments appearing in the heptagon orthoscheme volume formula of ref.~\cite{Ren:2023tuj}. No qualitatively new algebraic structures are generated by the recursion: the entire function space of $\cI_6^0$ is inherited from the heptagon at $\varepsilon = 0$ via the limiting procedure described above. The same statement is expected to hold for higher orders in $\varepsilon$ and higher multiplicities, with the function space at $\cO(\varepsilon^k)$ for the $N$-point integral inherited from the $(N+1+2k)$-point integer-dimensional integral.

Most of the involved polylogarithms are regular in the limit \(\cQ_{7,7} \to 0\). However, special care is required for terms with arguments of the form \(\sqrt{1- \frac{\cQ_{7,7}}{\cQ_{i,i}}}\), which may generate logarithmic divergences. 
These divergences are precisely compensated by the subtraction term in eq.~\eqref{eq:divergencies}, in agreement with the general structure of the recursion.

To isolate the singular contributions, we expand around \(\cQ_{7,7}=0\). Since only the logarithmic behaviour is relevant, we may replace
\beq
\sqrt{1- \frac{\cQ_{7,7}}{\cQ_{i,i}}} = 1 - \frac{\cQ_{7,7}}{2 \cQ_{i,i}} + \mathcal{O}(\cQ_{7,7}^2),
\eeq
which removes the square roots at leading order.\footnote{
This also removes the need for rationalising square roots, in contrast to the approach of ref.~\cite{Duhr:2025azh}.
}
This allows us to rewrite \(\mathcal{I}^{-1}_{7,8} (\check{\cQ})\) in a fibration basis with respect to \(\cQ_{7,7}\). In this basis, the result takes the form
\beq
2 \mathcal{I}^{-1}_{7,8} (\check{\cQ})
= - \mathcal{I}^{0}_{6} (\cQ)
- \log ( \cQ_{7,7} ) \,  \mathcal{I}^{-1}_{6} (\cQ)
+ \mathrm{o}(1),
\eeq
which is precisely the structure predicted by eq.~\eqref{eq:recursion}. 
This demonstrates that the heptagon integral in $D=8$ encodes both the $\eps^0$ and $\eps^1$ contributions of the hexagon.

At order $\mathcal{O}(\varepsilon)$, the subtraction term in eq.~\eqref{eq:divergencies} contributes only logarithmic divergences as $\cQ_{7,7} \to 0$ and therefore does not contribute to the finite part.
Therefore, extracting the finite part is equivalent to discarding the logarithmic divergence. In practice, 
we implement this by setting $\log(\cQ_{7,7}) \to 0$ at the level of the expansion of \(\mathcal{I}^{-1}_{7,8} (\check{\cQ})\), rendering the subtraction term irrelevant,
before taking the limit.
We stress that this simplification is specific to the $\mathcal{O}(\varepsilon)$ coefficient. At higher orders in \(\eps\), the subtraction term also contributes to the finite part, and a proper subtraction procedure becomes necessary. While such a subtraction can in principle be carried out systematically, it typically involves non-trivial relations among multiple polylogarithms, making it technically more involved. Nevertheless, performing the full subtraction provides a useful consistency check of the result obtained through the simplified procedure.

We implemented this computation in \texttt{Mathematica} using \texttt{PolyLogTools}~\cite{Duhr:2019tlz} and verified the result numerically against direct Feynman-parameter integration and sector decomposition using \texttt{pySecDec}~\cite{Borowka:2018goh,Heinrich:2021dbf,Heinrich:2023til}. Excellent agreement is observed at multiple points in Euclidean kinematics. 

This example provides a non-trivial validation of the recursion at higher multiplicity and demonstrates explicitly how the $\eps$-expansion is controlled by higher-point functions. Moreover, there is no conceptual obstruction to extending this procedure to higher multiplicities and higher orders in $\eps$. 

However, the rapid growth in complexity, already visible at the level of the hexagon at $\mathcal{O}(\eps)$, indicates that practical applications require further simplifications. It would therefore be interesting to investigate whether the resulting expressions can be systematically reduced using relations among multiple polylogarithms.

\section{Conclusions}
\label{sec:conclusions}

In this paper we have identified a previously unknown recursive structure in dimensionally regularised scalar one-loop Feynman integrals using Schläfli's differential equation. This recursive structure connects higher-order Laurent coefficients in the dimensional regulator \(\eps\) to special limits of Feynman integrals with a greater number of external legs. Since all integer dimensional integrals are known~\cite{Ren:2023tuj} in terms of MPLs through their connection to hyperbolic geometry~\cite{Davydychev:1997wa,Rudenko2020Orthoschemes}, our work provides an explicit recursive algorithm that allows one to compute any one-loop Feynman integral depending on any number of scales and external legs to any desired order of the dimensional regulator \(\eps\) in terms of MPLs. Taken together, our results demonstrate that the analytic structure of scalar one-loop integrals in Euclidean kinematics is governed by the hyperbolic geometry results.

The recursion relies on three main ingredients: the dissection of the Feynman integral into orthoschemes as in ref.~\cite{Duhr:2025azh}, the special representation of the dimensionally regularised integral as a one-fold integral over the \(\eps=0\) result, and Schl\"afli's differential equation~\cite{Schlafli1901}.

As a non-trivial application of the recursion, we computed the \(\mathcal{O}(\eps)\) term of the hexagon integral in \(D=6-2\eps\) dimensions, allowing for arbitrary internal masses and off-shell external momenta in Euclidean kinematics. As lower-point illustrations, in section~\ref{app:examples} we computed the bubble integral through \(\mathcal{O}(\eps^2)\), including a geometric interpretation of its \(\mathcal{O}(\eps)\) term, and worked out the box and triangle integrals at \(\mathcal{O}(\eps)\). For the triangle, we verified analytic agreement with the known result of ref.~\cite{Duhr:2025azh}, while for the box we found numerical agreement. In contrast to the representation of the box in ref.~\cite{Duhr:2025azh}, which involves nested square roots, our result contains only polylogarithms with at most a single square root in their arguments and no nested algebraic structures. The results obtained throughout this work have been checked against known analytic results and numerical evaluations where available.

While our results provide a recursive construction expressing all one-loop $N$-point integrals in terms of MPLs to all orders in $\varepsilon$, several natural extensions remain. Most importantly, it would be interesting to understand whether similar geometric recursions exist beyond one loop. Moreover, it would be desirable to clarify the algebraic structure underlying our geometrically motivated recursion — for example, whether antipodal duality~\cite{Arkani-Hamed:2017ahv} admits an extension to dimensional regularisation, or whether connections to cluster algebras can be established. Finally, at one loop, orthoscheme integrals appear to play a more fundamental role than the full Feynman integral itself. It remains an open question whether this is a general structural feature or a special property of the one-loop case.

\acknowledgments
I am grateful to Claude Duhr for supervision, numerous discussions, and comments on the manuscript. I would also like to thank Babis Anastasiou for discussions that initiated this line of research, and Herbert Gangl and Steven Charlton for helpful discussions. 

\appendix
\renewcommand{\appendixname}{}

\section{Even–odd leg conversion}\label{app:evenodd}
Our central result in eq.~\eqref{eq:recursion} expresses the \(\varepsilon^k\) contribution of an \(N\)-point function with \(N\) even through the  \(\varepsilon^{k-1}\) contribution of an \((N+1)\)-point function. 
To close the recursion, we therefore need a way to relate odd-point integrals back to even ones.
This relation is closely related to the fact that \((N-1)\)-point functions in \(N\) dimensions can be viewed as \(N\)-point functions in \(N\) dimensions with one leg at infinity. In the simplex picture this amounts to one vertex becoming ideal.

Proving relation \eqref{eq:eventoodd} can be done by using the Feynman parameter representation involving the Symanzik polynomial \(\mathcal{U}\) and \(\mathcal{F}\). 
The coefficients \(\mathcal{I}_N^k (\check{\cQ})\) and \(\mathcal{I}_{N-1,N}^k (\cQ)\) can be related to the Feynman parameter representation with Symanzik polynomials 
\begin{align}
    \mathcal{U} = \sum_{i=1}^{N-1} a_i  &\qquad \textnormal{and} \qquad
    \mathcal{F} = \sum_{i,j=1}^{N-1} \cQ_{i,j}\ a_i\, a_j,\\
    \check{\mathcal{U}} = \mathcal{U}+ a_N &\qquad \textnormal{and} \qquad \check{\mathcal{F}} = \mathcal{F} + \cQ_{N,N} a_N ( 2\, \mathcal{U} + a_N).
\end{align}
Then we can express the integrals through:
\begin{align}
    \sum_{-1 \le k \le \infty }\mathcal{I}_N^k (\check{\cQ}) \eps^k
    &= \int_{[0,\infty)^N} \mathrm{d}a\ \delta(1-\mathcal{U})
       \frac{ \Gamma(-\varepsilon) \Gamma\left(\frac{N}{2} + \varepsilon \right) }
            {\check{\mathcal{F}}^{\frac{N}{2}}}
       \left(\frac{\check{\mathcal{U}}^2}{\check{\mathcal{F}}} \right)^\varepsilon \, ,\\
    \sum_{-1 \le k \le \infty } \mathcal{I}_{N-1,N}^k (\cQ) \eps^k
    &= \int_{[0,\infty)^{N-1}} \mathrm{d}a\ \delta(1-\mathcal{U})
       \frac{ \Gamma(-\varepsilon) \Gamma\left(\frac{N}{2}-1 + \varepsilon \right) }
            {\mathcal{U}\,\mathcal{F}^{\frac{N}{2}-1} }
       \left(\frac{\mathcal{U}^2}{\mathcal{F}} \right)^\varepsilon \, ,
\end{align}

Here we used projective invariance to impose \(\mathcal U=1\). The resulting \((N-2)\)-dimensional simplex for \(a_1,\ldots,a_{N-1}\) is compact and, for generic Euclidean kinematics, there are constants \(0<F_{\min}\leq\mathcal F\leq F_{\max}<\infty\). Although \(\mathcal U=1\) on this gauge slice, we keep factors of \(\mathcal U\) explicit below to make the comparison of the two parameter representations transparent.
In the following we assume \(N\geq4\).  
To prove eq.~\eqref{eq:eventoodd}, set \(M\coloneqq\cQ_{N,N}\) and combine
the two powers in the first Feynman-parameter representation.  After
multiplication by \(2M\), the integral over \(a_N\) is
\begin{equation}
L_M(\eps)
\coloneqq
2M\int_0^\infty \mathrm da_N\,
(\mathcal U+a_N)^{2\eps}
\left[\mathcal F+M a_N(2\mathcal U+a_N)\right]^{-\frac N2-\eps}\, .
\end{equation}
Consequently, \(L_M\) is related to the full integral by the exact identity
\begin{equation}
 2M\sum_{k=-1}^{\infty}\mathcal I_N^k(\check{\cQ})\eps^k
 =\Gamma(-\eps)\Gamma\left(\frac N2+\eps\right)
 \int_{[0,\infty)^{N-1}}
 \left(\prod_{i=1}^{N-1}\mathrm da_i\right)
 \delta(1-\mathcal U)\,L_M(\eps)\, .
\end{equation}
With the change of variables \(t=2\mathcal U M a_N\), one obtains the exact identity
\begin{equation}
L_M(\eps)
=\frac{1}{\mathcal U}\int_0^\infty\mathrm dt\,
\left(\mathcal U+\frac{t}{2\mathcal U M}\right)^{2\eps}
\left(\mathcal F+t+\frac{t^2}{4\mathcal U^2M}\right)^{-\frac N2-\eps}\, .
\end{equation}
On the gauge slice \(\mathcal U=1\), let
\(r\coloneqq\operatorname{Re}\eps\) and
\(c\coloneqq\min\{1,F_{\min}\}>0\).  For \(M\geq1\),
\begin{equation}
 \mathcal F+t+\frac{t^2}{4M}\geq c(1+t),
 \qquad
 1+\frac{t}{2M}\leq1+t\, .
\end{equation}
It follows that, uniformly in \(M\) and the remaining Feynman parameters,
\begin{equation}
 \left|
 \left(1+\frac{t}{2M}\right)^{2\eps}
 \left(\mathcal F+t+\frac{t^2}{4M}\right)^{-N/2-\eps}
 \right|
 \leq C(1+t)^{-N/2+|r|}
 \leq C(1+t)^{-N/2+\eta}\, .
\end{equation}
for \(|r|\leq\eta<\frac N2-1\).  The last expression is integrable for
\(N\geq4\).
Dominated convergence therefore applies both to the \(t\)-integration and
to the compact simplex fixed by \(\mathcal U=1\), and gives locally
uniformly in \(\eps\)
\begin{align}
\lim_{M\to\infty}L_M(\eps)
&=\mathcal U^{2\eps-1}
\int_0^\infty\frac{\mathrm dt}{(\mathcal F+t)^{\frac N2+\eps}}\\
&=\frac{\mathcal U^{2\eps-1}
\mathcal F^{-\left(\frac N2-1+\eps\right)}}
{\frac N2-1+\eps}\\
&=\frac{1}{\left(\frac N2-1+\eps\right)
\mathcal U\,\mathcal F^{\frac N2-1}}
\left(\frac{\mathcal U^2}{\mathcal F}\right)^\eps\, .
\end{align}
Finally,
\begin{equation}
\frac{\Gamma\left(\frac N2+\eps\right)}
{\frac N2-1+\eps}
=\Gamma\left(\frac N2-1+\eps\right)\, .
\end{equation}
Substitution into the exact representation above shows that
\begin{equation}
 \lim_{M\to\infty}
 2M\sum_{k=-1}^{\infty}\mathcal I_N^k(\check{\cQ})\eps^k
 =\sum_{k=-1}^{\infty}\mathcal I_{N-1,N}^k(\cQ)\eps^k\, .
\end{equation}
This proves eq.~\eqref{eq:eventoodd}.

\bibliographystyle{JHEP}

\providecommand{\href}[2]{#2}\begingroup\raggedright\endgroup

\end{document}